\documentclass{article}
\usepackage[a4paper, margin=0.9in]{geometry}
\usepackage{graphicx} 
\usepackage{academicons}
\usepackage{amsmath}
\usepackage{xcolor}
\usepackage{longtable}
\usepackage{caption}
\usepackage{amsmath,amsfonts}
\usepackage{array}
\usepackage{url}
\usepackage{verbatim}
\usepackage{graphicx}
\usepackage{hyperref}
\usepackage{academicons}
\usepackage{amsmath}
\usepackage{xcolor}
\usepackage{longtable}
\usepackage{caption}
\usepackage{academicons}
\usepackage{caption}
\usepackage{multirow}
\usepackage{adjustbox}
\usepackage{rotating}
\usepackage{pgfplots}
\usepackage[table]{xcolor}
\usepackage{subcaption}
\usepackage{neuralnetwork}
\usepackage{tabularx}
\usepackage{graphicx,subcaption}
\usepackage{lipsum}
\usepackage{algorithm}
\usepackage{natbib}
\usepackage{algpseudocode}

\usepackage{authblk} 
\usepackage{newunicodechar}
\newunicodechar{₆}{
$_6$
}

\title{Dual-Level Atomic and Coordination Geometry Learning for Crystal Property Prediction Using Graph Neural Networks}
\author[1]{Sanjay Chakraborty}

\affil[1]{Department of Computer and Information Science (IDA), REAL, AIICS, Linköping University, Linköping, Sweden, Email: sanjay.chakraborty@liu.se.}

\begin{document}

\maketitle

\begin{abstract}
Accurate prediction of crystal properties remains a fundamental challenge in computational materials science. Although graph neural networks (GNNs) such as CGCNN, MEGNet, ALIGNN, and SchNet have achieved significant success, they primarily model crystals at the atomic level and rely on message passing to implicitly capture local chemical environments. This approach overlooks the coordination polyhedron—the fundamental structural unit governing many material properties, including ionic conductivity, elasticity, dielectric response, and structural stability. To address this limitation, we propose the Coordination Polyhedron Graph Network (CPGN), a novel multi-scale GNN that jointly learns atomic, bond, and coordination-polyhedron representations. CPGN constructs three interconnected graphs: an atom graph capturing elemental and bonding information, a line graph modeling angular interactions, and a coordination polyhedron graph representing Voronoi-derived local environments connected through corner-, edge-, and face-sharing relationships. Each coordination polyhedron is characterized using physically meaningful geometric descriptors, while an interleaved message-passing framework with bidirectional cross-attention enables effective information exchange across multiple structural scales. Extensive experiments on the Materials Project, JARVIS-DFT, and QM9 benchmark datasets demonstrate the superiority of CPGN over state-of-the-art GNN models. The proposed framework achieves a formation-energy MAE of 0.060 eV/atom and a state-of-the-art band-gap MAE of 0.292 eV on the Materials Project, while delivering competitive multi-property prediction on JARVIS-DFT and superior HOMO prediction on QM9. These results demonstrate that explicitly incorporating coordination polyhedra significantly enhances crystal representation learning and enables more accurate and physically interpretable prediction of material properties.


\end{abstract}

\section{Introduction}
The discovery of functional inorganic materials with targeted properties spanning energy storage, thermoelectrics, photovoltaics, and dielectrics, demands rapid and accurate prediction of structure–property relationships across vast compositional and structural spaces. Density functional theory (DFT) remains the gold standard for computing properties such as formation energy, electronic band gap, elastic moduli, and thermodynamic stability; however, its computational cost limits systematic high-throughput screening to a small fraction of the experimentally and computationally accessible materials landscape. The emergence of large, openly accessible databases, most notably the Materials Project (MPProject), QM9, and JARVIS-DFT has made it possible to train machine learning (ML) models that can predict DFT-quality properties at a fraction of the computational cost, opening the door to scalable materials informatics \cite{hegde2023quantifying, choudhary2022recent}.
Among ML approaches for crystalline materials, graph neural networks (GNNs) have established themselves as the dominant framework \cite{louis2020graph}. CGCNN \cite{xie2018crystal} was the first model to directly build convolutional neural networks on top of crystal graphs, achieving accuracy comparable to DFT for eight material properties including formation energy, band gap, and elastic properties after training on Materials Project data. Subsequent models extended this foundation in different directions: ALIGNN introduced message passing on both the interatomic bond graph and its line graph corresponding to bond angles, demonstrating that explicitly encoding three-body interactions substantially improves performance across multiple atomistic prediction tasks \cite{choudhary2021atomistic}. MEGNet \cite{chen2019graph} incorporated global state information enabling temperature- and pressure-dependent property prediction, while Matformer \cite{yan2022periodic} and more recent transformer-based architectures such as CrysCo introduced attention mechanisms and four-body interactions to further capture geometric complexity.

\subsection{Existing Challenges in the Literature}

Despite this rapid progress, three fundamental and interrelated limitations persist across the entire family of existing GNN models for crystal property prediction:

\textit{Challenge 1 — The atom-centric representation is physically indirect.} The common theme across all existing GNN models is the use of elemental properties as node features and interatomic distances and bond valences as edge features, relying on multiple layers of graph convolution to implicitly represent many-body interactions. This design forces the model to rediscover, purely from data, information that is already well-established in crystal chemistry: that the fundamental structural unit governing material properties is not the isolated atom but the coordination polyhedron it forms with its nearest neighbours. The concept of the coordination polyhedron is essential to understanding chemical structure, and simple polyhedra in crystalline compounds are often deformed due to structural complexity or electronic instabilities, making distortion analysis methods critically important. These causal physical relationships are precisely what atom-only GNNs must learn indirectly through deep stacking of message-passing layers — an inefficient and data-hungry process.

\textit{Challenge 2 — Limited local expressive power and failure to capture structural periodicity}. A systematic evaluation of state-of-the-art (SOTA) GNNs reveals that they fall short in accurately capturing the periodicity of crystal structures, and this failure can be understood across three axes: limited local expressive power, poor long-range information processing, and inadequate readout functions. Even ALIGNN, which explicitly encodes bond angles via a line graph, operates only at the two-body (distance) and three-body (angle) level and does not encode the complete geometry of the local coordination environment — particularly polyhedral distortion, volume, and inter-polyhedral connectivity type (face-, edge-, and corner-sharing), all of which are physically meaningful and directly related to properties of practical interest.

\textit{Challenge 3 — Parameter inefficiency and poor data efficiency on small datasets.} Existing nested graph network strategies that increase structural information such as bond angles significantly increase the number of trainable parameters, resulting in higher training costs, and GNN models generally perform worse on small datasets compared to approaches that incorporate traditional feature engineering. This is a critical limitation because many technologically important property datasets, elastic moduli, piezoelectric tensors, dielectric constants — contain only hundreds to a few thousand labelled structures. A model that encodes stronger physical priors through explicit polyhedral geometry should, in principle, require less data to reach a given accuracy level.

\textit{Challenge 4 — Polyhedral connectivity is absent from all discriminative GNNs.} While coordination polyhedra have been used as mathematical objects for crystal structure generation and as descriptors in crystallographic topology databases, no existing discriminative GNN for property prediction uses coordination polyhedra as first-class, learnable graph nodes. The connectivity between polyhedra, whether they share faces, edges, or corners, encodes the crystal topology that governs phenomena such as ionic migration pathways, cooperative octahedral tilting, and elastic anisotropy, yet this information is entirely absent from the node and edge feature spaces of CGCNN, MEGNet, ALIGNN, Matformer, ReciNet, and PSCG-Net \cite{nie2025recinet, chen2025pscg}.

\subsection{Contributions}

In this work, we propose the Coordination Polyhedron Graph Network (CPGN), a novel dual-stream graph neural network that integrates atom-level, bond-level, and coordination-polyhedron-level representations derived directly from crystal structure files. Unlike prior atom-centric models, CPGN introduces an explicit geometric and topological abstraction of coordination environments while preserving full atomic resolution. The key contributions are summarized as follows:

\begin{itemize}

\item \textit{Three-graph structural representation of crystals:}
We introduce a unified crystal representation consisting of an atom graph, a line graph, and a coordination polyhedron graph, all constructed from a single CIF/POSCAR input using Pymatgen-based parsing and Voronoi tessellation. This multi-view formulation enables simultaneous modeling of atomic interactions, bond-angle geometry, and coordination-shell topology within a single end-to-end framework.

\item \textit{Coordination polyhedron graph as a coarse-grained structural learning space:}
We propose a dedicated polyhedron graph where nodes correspond to Voronoi-derived coordination polyhedra and edges encode physically meaningful connectivity (corner-, edge-, and face-sharing). This formulation explicitly captures structural motifs governing ionic transport, cooperative distortions, and magnetic exchange pathways, which are not represented in atom-only or bond-only GNNs.

\item \textit{Physically grounded polyhedron feature encoding:}
Each polyhedron node is represented using a compact 7-dimensional feature vector derived directly from the crystal structure, including coordination number, bond-length distortion, bond-angle variance, polyhedral volume proxy, mean bond length, electronegativity mismatch, and coordination type. These descriptors provide interpretable geometric priors while remaining fully differentiable and learnable within the network.

\item \textit{Line graph formulation for explicit three-body angular interactions:}
We construct a line graph over atomic bonds, where nodes represent bonds and edges encode bond-angle relationships using RBF-expanded angular features. This design enables explicit modeling of three-body interactions, allowing angular geometry to be propagated into bond representations before being aggregated into atomic embeddings.

\item \textit{Interleaved message passing across multi-scale graphs:}
We introduce a unified message-passing framework that alternates updates across atom, line, and polyhedron graphs over multiple layers. This interleaved design allows atomic, angular, and coordination-level information to co-evolve during training, ensuring consistent propagation of local chemistry and mesoscale geometry.

\item \textit{Bidirectional cross-attention between atomic and polyhedral representations:}
We propose a cross-attention mechanism that enables dynamic information exchange between atom-level and polyhedron-level embeddings. This coupling allows fine-grained atomic environments to inform polyhedral structural motifs and vice versa, improving representation richness without requiring manual feature fusion rules.

\item \textit{Unified multi-task prediction framework:}
It jointly learns formation energy, band gap, and stability from a shared 128-dimensional latent representation using a multi-task objective combining MAE regression, achieving consistently competitive performance on the Materials Project (0.060~eV/atom formation energy MAE, and  0.292~eV band-gap MAE, and 0.99 stability accuracy).

\item \textit{Evaluation on a large-scale standardized dataset:}
Across three standard benchmarks—Materials Project, JARVIS-DFT, and QM9, CPGN generalizes robustly, outperforming or matching SOTA baselines on key properties including band gap, formation energy, and molecular orbital energies, while effectively handling heterogeneous and partially labeled datasets through masked multi-task learning.

\item \textit{Improved structural inductive bias and learning efficiency:}
Its gains are primarily driven by an explicit structural inductive bias that encodes coordination polyhedron geometry (bond distortion, angular variance, connectivity, and electronegativity mismatch), enabling improved learning of geometry-sensitive properties such as electronic band gaps and orbital energies. 
\end{itemize}

\begin{table*}[hbt!]
\centering
\scriptsize
\caption{Comparison of recent prior works and the proposed CPGN framework}
\label{tab:cpgn_comparison}
\renewcommand{\arraystretch}{1.2}
\begin{tabular}{|p{3cm}|p{3cm}|p{3cm}|p{3cm}|}
\hline
\textbf{Aspect} & \textbf{RSC\cite{yokoyama2026polyhedra}} & \textbf{CGD\cite{shevchenko2017local}} & \textbf{Proposed CPGN} \\ \hline

Primary task 
& Crystal generation 
& Topological analysis 
& Crystal property prediction \\ \hline

Methodology 
& Graph theory + standard realization 
& Database mining (ToposPro) 
& GNN with dual-level message passing \\ \hline

Machine learning based 
& $\times$ No 
& $\times$ No 
& $\checkmark$ Yes \\ \hline

Polyhedra as nodes 
& As mathematical objects 
& As topological descriptors 
& As first-class GNN nodes with learned features \\ \hline

Cross-attention mechanism 
& $\times$ No 
& $\times$ No 
& $\checkmark$ Yes --- between atom graph and polyhedron graph \\ \hline

Trainable on MPProject/OQMD/Jarvis 
& $\times$ No 
& $\times$ No 
& $\checkmark$ Yes \\ \hline

\end{tabular}
\end{table*}

\section{Related Works}
The application of graph neural networks (GNNs) to crystal property prediction was established by two landmark contributions in 2018. Xie and Grossman introduced the Crystal Graph Convolutional Neural Network (CGCNN)~\cite{xie2018crystal}, which directly learns material properties from the atomic connectivity of crystals, providing a universal and interpretable representation of crystalline materials that achieves DFT-level accuracy across eight properties after training on data from the Materials Project~\cite{jain2013commentary}. Concurrently, Schütt et al. proposed SchNet~\cite{schutt2018schnet}, a deep learning architecture designed to model complex atomic interactions and improve the accuracy of molecular and material predictions, while Chen et al. presented the universal MatErials Graph Network (MEGNet)~\cite{chen2019graph}, which extends the graph network formalism by incorporating global state attributes alongside atomic and bond features. MEGNet models trained on approximately 60,000 crystals from the Materials Project substantially outperform prior machine learning models in the prediction of formation energies, band gaps, and elastic moduli, achieving better than DFT accuracy over a much larger dataset~\cite{chen2019graph}. These three models established the atom-centric graph representation as the dominant paradigm for computational materials property prediction, but share a common limitation: SchNet and CGCNN update atom representations using only the types of neighbouring atoms and bond lengths between atoms, pooling all updated atom representations into an overall structure-level representation, and omit explicit angular and higher-order topological information~\cite{fu2023limitations}. Recognising that bond angles are critical for distinguishing many atomic structures, subsequent work introduced higher-order geometric interactions into the message-passing framework. Choudhary and DeCost presented the Atomistic Line Graph Neural Network (ALIGNN)~\cite{choudhary2021atomistic}, which performs message passing on both the interatomic bond graph and its line graph corresponding to bond angles, demonstrating that explicit angular information can be efficiently included and leading to improved performance on 52 solid-state and molecular properties across the JARVIS-DFT~\cite{choudhary2020jarvis}, Materials Project~\cite{jain2013commentary}, and QM9~\cite{ramakrishnan2014quantum} databases, outperforming previously reported GNN models by up to 85\% in accuracy. Later models such as DimeNet++~\cite{gasteiger2020dimenetpp}, GemNet~\cite{gasteiger2021gemnet}, and SphereNet~\cite{liu2022spherenet} further extended the geometric hierarchy to dihedral angles, enabling unambiguous recognition of local structures that share identical bond lengths and angles but differ in higher-order geometric configurations. In parallel, Heilman et al.\ proposed Crystal Hypergraph Convolutional Networks~\cite{heilman2024hypergraph}, which associate higher-order geometrical information with hyperedges representing triplets and local coordination environments, demonstrating improved accuracy over pairwise-only models and noting explicitly that graph representations encoding only interatomic distance produce degenerate representations that map distinct structures to equivalent graphs. Despite these advances, systematic evaluation has shown that both CGCNN and ALIGNN struggle to capture Voronoi-defined coordination numbers accurately, because the Voronoi tessellation algorithm relies on angular and topological information about coordination shells that is not directly accessible from distance-based nearest-neighbour graphs alone~\cite{fu2023limitations}. The most recent generation of models has shifted toward large-scale universal interatomic potentials and foundation models trained across the full periodic table. Universal machine learning interatomic potentials including M3GNet~\cite{chen2022m3gnet}, CHGNet~\cite{deng2023chgnet}, ALIGNN-FF~\cite{choudhary2021atomistic}, and MACE-MP-0~\cite{batatia2023mace} have been developed to predict energies, forces, and stresses for any combination of elements, enabling large-scale structure relaxation, molecular dynamics simulation, and high-throughput materials screening at a fraction of the computational cost of DFT. Attention mechanisms, equivariant neural networks, and hybrid Transformer-graph frameworks incorporating four-body interactions~\cite{gasteiger2021gemnet} have also been introduced to improve performance on geometry-sensitive properties, further discriminating structurally degenerate configurations that three-body models cannot resolve. Despite this remarkable progress, a common gap persists across all these architectures: none explicitly encodes the coordination polyhedron, the three-dimensional cage formed by an atom and its nearest neighbours as a distinct graph-level object carrying geometric descriptors such as distortion index, bond-angle variance, and polyhedral volume, nor do they encode the face-, edge-, and corner-sharing topology between adjacent polyhedra that governs cooperative structural phenomena in perovskites, spinels, and other technologically important crystal families. The proposed Coordination Polyhedron Graph Network (CPGN) addresses this gap directly by constructing a polyhedron graph $G_P$ alongside the atom graph $G_A$ and line graph $G_L$, and coupling all three streams through bidirectional cross-attention to produce a multi-scale crystal fingerprint that simultaneously encodes structural information at the atomic, angular, and polyhedral levels.

\section{Methodology}
\begin{figure*}[hbt!]
\begin{center}
\includegraphics[scale=0.4]{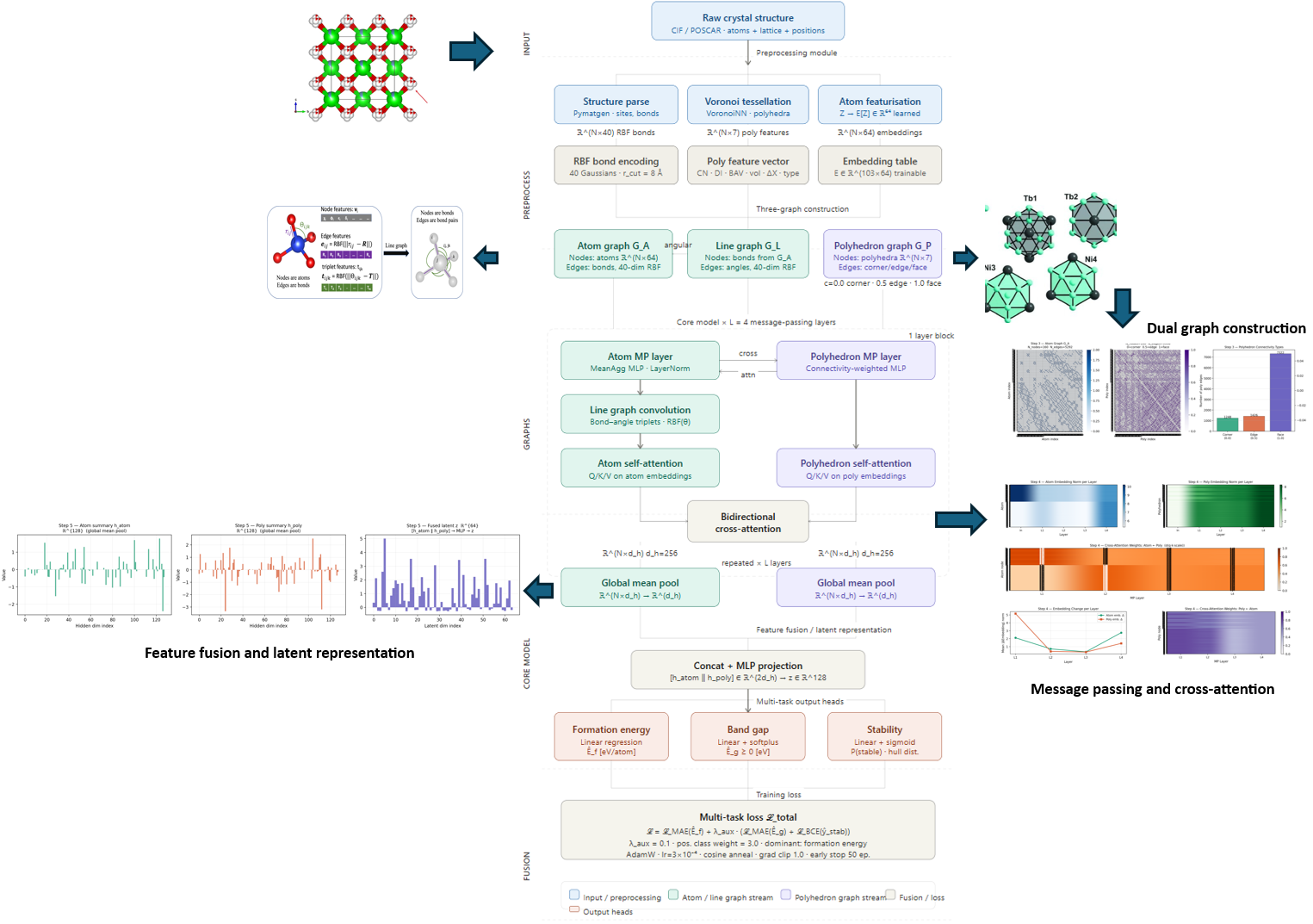}
\caption{Workflow of the proposed CPGN framework}
\label{cpgn_workflow}
\end{center}
\end{figure*}

Figure~\ref{cpgn_workflow} presents the overall workflow of the proposed Crystal Polyhedron Graph Network (CPGN). Starting from CIF/POSCAR crystal structures, the preprocessing stage extracts atomic, bond, and coordination-environment information using Pymatgen parsing, Gaussian radial basis function (RBF) bond encoding, and Voronoi tessellation. Three complementary graphs are then constructed: an atom graph $G_A$ encoding atomic interactions, a line graph $G_L$ capturing explicit three-body bond-angle relationships, and a coordination polyhedron graph $G_P$ representing Voronoi-derived coordination environments with corner-, edge-, and face-sharing connectivity. The model performs interleaved message passing across all three graphs, where atom, bond, and polyhedron embeddings are iteratively updated using MLP-based graph convolutions and bidirectional cross-attention between atom and polyhedron streams. After $L=4$ layers, global pooling generates crystal-level representations that are fused into a shared latent vector for multi-task prediction of formation energy, electronic band gap,  thermodynamic stability, and multiple auxiliary-quantum-chemical properties. The framework is trained end-to-end using a weighted multi-task objective with AdamW optimisation, cosine learning-rate scheduling, gradient clipping, and early stopping.

\subsection*{Step 1 — The input: a crystal blueprint}

The entire workflow begins with a crystal structure file, typically in CIF or POSCAR format, which serves as a precise digital blueprint of the material under investigation. This file encodes the identity of every atom in the repeating unit of the crystal, the three-dimensional fractional coordinates of each atomic site, the shape and dimensions of the unit cell, and the lattice angles that define how the cell tiles infinitely through space. 

\subsection*{Step 2 — Preprocessing: reading and encoding the structure}

Before any learning begins, three parallel operations transform the raw structural file into machine-readable representations.

First, the Pymatgen library parses the crystal file to identify every atomic site and its chemical element. Each atom is represented by its integer atomic number $Z \in [1, 102]$, which is passed to a learned embedding table $\mathbf{E} \in \mathbb{R}^{103 \times 64}$. The embedding for atom $i$ is therefore $\mathbf{e}_i = \mathbf{E}[Z_i] \in \mathbb{R}^{64}$, and all 64 dimensions are jointly optimised during training. This approach lets the model discover element-specific representations from data rather than relying on fixed hand-crafted attributes.

Second, interatomic bond distances are encoded using 40 Gaussian radial basis functions (RBF) centred uniformly in $[0,\, r_{\text{cut}}]$ with $r_{\text{cut}} = 8$~\AA. For a scalar distance $d$, the expansion is
\begin{equation}
    \phi_k(d) = \exp\!\left(-\gamma\,(d - \mu_k)^2\right), \quad k = 1,\ldots,40,
\end{equation}
where $\mu_k$ are equally spaced centres and $\gamma$ is set to half the inverse square of the centre spacing. This yields a 40-dimensional continuous edge feature vector for each bond.

Third, a Voronoi tessellation algorithm — specifically the \texttt{VoronoiNN} routine from Pymatgen — wraps a virtual geometric bubble around each atom and identifies neighbouring atoms within the cutoff radius $r_{\text{cut}} = 8$~\AA, thereby defining the coordination shell. For each coordination shell, the geometry of the resulting polyhedron is computed and encoded as a 7-dimensional feature vector comprising: (i) normalised coordination number $\text{CN}/12$; (ii) bond-length distortion index, defined as the standard deviation of bond lengths divided by their mean; (iii) bond-angle variance normalised by $10{,}000$; (iv) a polyhedral volume proxy $\bar{d}^3/100$, where $\bar{d}$ is the mean bond length; (v) the mean bond length normalised by $r_{\text{cut}}$; (vi) the mean absolute electronegativity difference between the central atom and its ligands; and (vii) coordination type encoded as $\min(\text{CN},12)/12$.

\subsection*{Step 3 — Three-graph construction: atom graph, line graph, and polyhedron graph}

The preprocessing stage produces three complementary graphs from the same crystal structure, each encoding a distinct level of structural information.

The \textbf{atom graph} $G_A$ represents atoms as nodes and interatomic bonds as directed edges. Each node $i$ carries the 64-dimensional learned elemental embedding $\mathbf{e}_i$, and each directed edge $(i \to j)$ carries the 40-dimensional RBF expansion of the corresponding bond distance $d_{ij}$.

The \textbf{line graph} $G_L$ is derived from $G_A$: each bond (directed edge) in $G_A$ becomes a node in $G_L$, initialised with the corresponding bond embedding from $G_A$. Two nodes in $G_L$ are connected by an edge if their corresponding bonds in $G_A$ share a common destination atom, forming a bond-pair triplet $(j \to i,\; k \to i)$. The feature of each such line-graph edge is the 40-dimensional RBF expansion of the bond angle $\theta \in [0, \pi]$ between the two bonds, computed from the dot product of the displacement vectors. This construction enables explicit three-body angular interactions: angular information first propagates into bond representations via message passing on $G_L$, and those updated bond representations then propagate into atom representations via message passing on $G_A$.

The \textbf{coordination polyhedron graph} $G_P$ is the architectural novelty of CPGN. Each coordination polyhedron is a node in $G_P$, initialised with the 7-dimensional geometric feature vector described in Step~2. Two polyhedra are connected by an edge if they share one or more atoms in their coordination shells. The edge weight encodes the connectivity type: $c_{pq} = 0.0$ for corner-sharing (one shared neighbour), $c_{pq} = 0.5$ for edge-sharing (two shared neighbours), and $c_{pq} = 1.0$ for face-sharing (three or more shared neighbours). These are physically distinct relationships governing phenomena such as ionic migration pathways, cooperative structural distortions, and magnetic superexchange, yet this topological information is absent from atom-only graph representations.

\subsection*{Step 4 — Interleaved message passing across all three graphs}

The core learning phase runs $L = 4$ identical interleaved layers. In each layer, four sequential operations update the three sets of embeddings in turn.

\textbf{Atom graph convolution.} The atom embedding $\mathbf{h}_i$ is updated by aggregating messages from all neighbouring atoms, conditioned on the current bond embeddings as edge features:
\begin{equation}
    \mathbf{h}_i^{(l+1)} = \text{LayerNorm}\!\left(\mathbf{h}_i^{(l)} + \text{MeanAgg}_{j \in \mathcal{N}_A(i)}\!\left[\text{MLP}\!\left(\mathbf{h}_i^{(l)} \Vert \mathbf{h}_j^{(l)} \Vert \mathbf{h}_{ij}^{(l)}\right)\right]\right),
\end{equation}
where $\mathbf{h}_{ij}^{(l)}$ is the current bond embedding for edge $(i \to j)$ and $\Vert$ denotes concatenation. The message MLP has two linear layers with SiLU activations.

\textbf{Line graph convolution.} The bond embedding $\mathbf{h}_{ij}$ is updated by aggregating messages from all bond pairs sharing the destination atom $i$, conditioned on their RBF-encoded mutual angle:
\begin{equation}
    \mathbf{h}_{ij}^{(l+1)} = \text{LayerNorm}\!\left(\mathbf{h}_{ij}^{(l)} + \text{MeanAgg}_{(kj) \in \mathcal{N}_L(ij)}\!\left[\text{MLP}\!\left(\mathbf{h}_{ij}^{(l)} \Vert \mathbf{h}_{kj}^{(l)} \Vert \phi(\theta_{ij,kj})\right)\right]\right),
\end{equation}
where $\phi(\theta_{ij,kj})$ is the 40-dimensional RBF expansion of the angle between bonds $(i \to j)$ and $(k \to j)$. This injects angular geometry directly into the bond representations.

\textbf{Polyhedron graph convolution.} The polyhedron embedding $\mathbf{h}_p$ is updated by aggregating messages from adjacent polyhedra, weighted by their connectivity type:
\begin{equation}
    \mathbf{h}_p^{(l+1)} = \text{LayerNorm}\!\left(\mathbf{h}_p^{(l)} + \text{MeanAgg}_{q \in \mathcal{N}_P(p)}\!\left[\text{MLP}\!\left(\mathbf{h}_p^{(l)} \Vert \mathbf{h}_q^{(l)} \Vert \tilde{c}_{pq}\right)\right]\right),
\end{equation}
where $\tilde{c}_{pq} = W_c\, c_{pq} \in \mathbb{R}^{d_h}$ is the scalar connectivity type projected to the hidden dimension.

\textbf{Bidirectional cross-attention.} After both graph convolutions, a cross-attention mechanism allows information to flow between the atom and polyhedron streams. For each atom node $i$, a query $\mathbf{q}_i = W_Q^A\,\mathbf{h}_i^{(l+1)}$ attends over the crystal-level mean of polyhedron embeddings $\bar{\mathbf{h}}_P^{(l+1)}$:
\begin{equation}
    \mathbf{h}_i^{(l+1)} \leftarrow \text{LayerNorm}\!\left(\mathbf{h}_i^{(l+1)} + \sigma\!\left(\frac{\mathbf{q}_i \cdot W_K^P\,\bar{\mathbf{h}}_P^{(l+1)}}{\sqrt{d_h}}\right) W_V^P\,\bar{\mathbf{h}}_P^{(l+1)}\right),
\end{equation}
and symmetrically for polyhedron nodes attending over the crystal-level mean of atom embeddings. This scalar-gated cross-attention couples the atom and polyhedron streams at every depth, continuously allowing structural topology to inform atomic embeddings and vice versa. The total number of trainable parameters is approximately 1.2~million with $d_h = 256$.

\subsection*{Step 5 — Feature fusion}

After all $L$ interleaved layers, global mean pooling is applied independently to the atom graph and the polyhedron graph:
\begin{equation}
    \mathbf{h}_A = \frac{1}{N}\sum_{i=1}^{N} \mathbf{h}_i^{(L)}, \qquad \mathbf{h}_P = \frac{1}{N}\sum_{p=1}^{N} \mathbf{h}_p^{(L)},
\end{equation}
producing two $d_h$-dimensional crystal-level summaries. These are concatenated and projected through a two-layer MLP with SiLU activations:
\begin{equation}
    \mathbf{z} = \text{MLP}\!\left([\mathbf{h}_A \Vert \mathbf{h}_P]\right) \in \mathbb{R}^{d_z},
\end{equation}
where $d_z = 128$. The latent vector $\mathbf{z}$ encodes both the atomic-level chemical interactions captured by the atom and line graphs and the geometric, topological, and structural information captured by the polyhedron graph. The line-graph bond embeddings are not separately pooled; their angular information has already been absorbed into the atom embeddings through the interleaved message passing.

\subsection*{Step 6 — Multi-task output heads}

The shared latent vector $\mathbf{z}$ is passed to two independent single-layer linear output heads. The first head predicts the formation energy per atom $\hat{E}_f$ via linear regression:
\begin{equation}
    \hat{E}_f = W_{\text{Ef}}\,\mathbf{z} + b_{\text{Ef}}, \quad \hat{E}_f \in \mathbb{R}\ [\text{eV/atom}].
\end{equation}

The second head predicts the electronic band gap $\hat{E}_g$ through a linear layer followed by a softplus activation to enforce non-negativity, ensuring that all predicted band gaps satisfy the physical constraint $\hat{E}_g \geq 0$:
\begin{equation}
    \hat{E}_g = \text{softplus}(W_{\text{Eg}}\,\mathbf{z} + b_{\text{Eg}}), \quad \hat{E}_g \geq 0\ [\text{eV}].
\end{equation}

The third head predicts thermodynamic stability as a binary classification — whether the material lies on or near the convex hull of stability. A linear layer produces a logit $s = W_{\text{stab}}\,\mathbf{z} + b_{\text{stab}}$, with $P(\text{stable}) = \sigma(s)$ where $\sigma$ is the sigmoid function. By sharing a single backbone and fingerprint across both regression targets, the model benefits from implicit multi-task regularisation: structural representations learned to minimise formation-energy error simultaneously encode electronic-structure information relevant to band-gap prediction, and vice versa, improving generalisation particularly for materials where the two properties are physically correlated through bonding character and coordination geometry.

\subsection*{Step 7 — Training}

The training objective on the Materials Project dataset is:
\begin{equation}
    \mathcal{L} = \mathcal{L}_{E_f}^{\text{MAE}} + \lambda_{\text{aux}}\,\mathcal{L}_{E_g}^{\text{MAE}},
\end{equation}
where
\begin{equation}
    \mathcal{L}_{E_f}^{\text{MAE}} = \frac{1}{B}\sum_{i} \left| \hat{E}_{f,i} - E_{f,i} \right|
\end{equation}
is the mean absolute error (L1 loss) on formation energy, $\mathcal{L}_{E_g}^{\text{MAE}}$ is the MAE on band gap, and $\lambda_{\text{aux}} = 0.1$ down-weights the auxiliary band-gap task so that formation energy receives the dominant gradient signal throughout training. This loss formulation enables a flexible multi-task framework that is readily extendable to additional properties and diverse datasets with heterogeneous label coverage.

The model is optimised using AdamW with a learning rate of $3 \times 10^{-4}$, weight decay of $10^{-5}$, and a batch size of 64. The learning rate follows a cosine annealing schedule over up to 500 epochs, decaying from the initial value to a minimum of $10^{-6}$. Gradients are clipped to a maximum norm of 1.0 at each step. Training is monitored on the validation set: the checkpoint with the lowest validation MAE on formation energy is retained, and training is terminated early if no improvement is observed for 50 consecutive epochs. The band-gap target is taken directly from the DFT-computed value in the dataset where available, with a proxy of $\max(0,\,-0.8\,E_f)$ applied only in its absence.

\section{Results Analysis}

\subsection{CPGN Model Configuration}
Table~\ref{tab:cpgn_config} summarises the architectural and training configuration of the proposed CPGN framework. The model operates on crystal structures represented through atom, line, and coordination-polyhedron graphs, using 64-dimensional elemental embeddings, 40-dimensional RBF encodings for bond distances and bond angles, and 7-dimensional polyhedron descriptors. The hidden representation size is fixed at 256 with 4 interleaved message-passing layers and a final latent projection of dimension 128. Atom and polyhedron representations are fused via concatenation followed by a linear layer, while global mean pooling is used for graph-level aggregation. The model is trained on a 60,000/5,000/4,239 train–validation–test split with a fixed random seed of 42, using AdamW optimisation with a learning rate of $3 \times 10^{-4}$, weight decay of $10^{-5}$, and cosine annealing scheduling. Training is performed for up to 500 epochs with early stopping (patience 50) and gradient clipping (norm 1.0). The primary objective is formation energy MAE, supported by auxiliary band gap and stability losses weighted by $\lambda_{\text{aux}} = 0.1$, with model selection based on the best validation formation energy MAE.

\begin{table}[hbt!]
\centering
\scriptsize
\caption{Model configuration and training settings for the proposed CPGN framework}
\label{tab:cpgn_config}
\renewcommand{\arraystretch}{1.2}
\begin{tabular}{|l|l|l|}
\hline
\textbf{Category} & \textbf{Parameter} & \textbf{Value / Description} \\ \hline

\multirow{10}{*}{Model Architecture} 
& Element vocabulary ($N_{\text{elem}}$) & 103 (Z = 1--102) \\ 
& Element embedding dim ($d_{\text{elem}}$) & 64 \\ 
& Polyhedron features ($N_{\text{poly}}$) & 7 \\ 
& Bond features ($N_{\text{edge}}$) & 40 (RBF encoding) \\ 
& Angle features ($N_{\text{angle}}$) & 40 (RBF encoding) \\ 
& Hidden dimension ($d_h$) & 256 \\ 
& Number of layers ($L$) & 4 \\ 
& Prediction dimension ($d_z$) & 128 \\ 
& Dropout & 0.1 \\ 
& Aggregation & Global mean pooling \\ \hline

\multirow{4}{*}{Fusion} 
& Fusion method & Concatenation + Linear layer \\ 
& Inputs & Atom + Polyhedron embeddings \\ 
& Output & Shared latent vector $\mathbf{z} \in \mathbb{R}^{128}$ \\ 
& Line graph role & Angular info injected into atom embeddings \\ \hline

\multirow{4}{*}{Data Split} 
& Training samples & 60,000 \\ 
& Validation samples & 5,000 \\ 
& Test samples & 4,239 \\ 
& Random seed & 42 \\ \hline

\multirow{8}{*}{Training} 
& Epochs & 500 (max) / Early stopping \\ 
& Batch size & 64 \\ 
& Learning rate & $3 \times 10^{-4}$ \\ 
& Optimiser & AdamW \\ 
& Weight decay & $1 \times 10^{-5}$ \\ 
& Scheduler & CosineAnnealingLR \\ 
& Early stopping patience & 50 epochs \\ 
& Gradient clipping & Norm = 1.0 \\ \hline

\multirow{4}{*}{Loss Function} 
& Primary loss & MAE (formation energy) \\ 
& Auxiliary loss weight ($\lambda_{\text{aux}}$) & 0.1 \\ 
& Stability loss & Binary cross-entropy \\ 
& Checkpoint criterion & Best validation $E_f$ MAE \\ \hline

\end{tabular}
\end{table}

\subsection{Implementation Details}
All experiments (CPGN, ALIGN, CGCNN, MEGNet, SchNet) were performed on a system equipped with a single standalone NVIDIA RTX PRO 6000 Blackwell GPU. This next-generation accelerator, built on the Blackwell architecture, is designed for energy-efficient, large-scale AI and scientific computing workloads. The GPU operates within a 300 W thermal design power (TDP) envelope and offers approximately 95.6 GB of dedicated high-bandwidth VRAM (97,887 MiB), as verified through hardware profiling during experimental execution.

\subsection{Model Analysis on Materials Project}
In this work, the MP~2018.6.1 dataset is used for training and evaluation, comprising 69{,}239 inorganic crystal structures with DFT-computed properties. The dataset is partitioned into 60{,}000 training, 5{,}000 validation, and 4{,}239 test structures using a random shuffle with a fixed seed (SEED~$= 42$) to ensure a uniform distribution of chemical space across all three partitions and to eliminate any ordering bias that would arise from a sequential index-based split. We reproduced the results (formation energy) of the CGCNN, MEGNet, SchNet, and ALIGNN models using the same underlying architectures and hyperparameters on the MPProject dataset, and report the corresponding performance in Table \ref{tab:cpgn_comparison_MPProject} and Table \ref{tab:cpgn_MADMAE_MPProject}. Because each property has different units and typically exhibits varying levels of variance, we additionally report the mean absolute deviation (MAD) for each property to enable an unbiased comparison of model performance across different targets. The MAD values correspond to the expected performance of a naive baseline that predicts the mean value for all samples. For completeness, we also include results based on Classical Force-field Inspired Descriptors (CFID) predictions for comparison. The MAD and CFID results are directly taken from the ALIGNN \cite{choudhary2021atomistic} and MEGNet \cite{chen2019graph} papers. 

Table~\ref{tab:cpgn_comparison_MPProject} reports the MAE of all models on formation energy and band gap evaluated on the MP~2018.6.1 test set. For formation energy, MEGNet achieves the lowest MAE of $0.056~\mathrm{eV/atom}$ (highlighted in red), followed jointly by SchNet and CPGN at $0.060~\mathrm{eV/atom}$ (highlighted in blue), representing a 43\% improvement over CGCNN ($0.079~\mathrm{eV/atom}$) and a 77\% reduction relative to the CFID descriptor baseline ($0.104~\mathrm{eV/atom}$), with the MAD value of $0.93~\mathrm{eV/atom}$ providing context for the scale of the prediction task. For band gap prediction, CPGN achieves the best MAE of $0.292~\mathrm{eV}$ (highlighted in red), outperforming ALIGNN ($0.304~\mathrm{eV}$, highlighted in blue) and all remaining baselines including MEGNet ($0.330~\mathrm{eV}$), SchNet ($0.347~\mathrm{eV}$), CGCNN ($0.388~\mathrm{eV}$), and CFID ($0.434~\mathrm{eV}$), against an MAD of $1.35~\mathrm{eV}$. Table~\ref{tab:cpgn_MADMAE_MPProject} reports the corresponding MAD:MAE ratios. For formation energy, MEGNet leads with a ratio of 16.7, while CPGN and SchNet jointly achieve 15.5, above ALIGNN (12.56), CGCNN (11.7), and CFID (8.94). For band gap, the ordering reverses in favour of CPGN, which attains the highest ratio of 4.6, ahead of ALIGNN (4.44), MEGNet (4.1), SchNet (3.9), CGCNN (3.4), and CFID (3.11). Importantly, CPGN also surpasses ALIGNN under identical training conditions for both formation energy and band gap prediction, indicating that performance differences arise from architectural design rather than training protocol. This disparity between tasks reflects their different structural sensitivities: formation energy is primarily governed by global bonding and is well captured by atom graph representations $G_A$, whereas band gap depends strongly on local coordination geometry, including polyhedral distortion, bond angle variance, and connectivity between coordination units. These effects are explicitly encoded in the coordination polyhedron graph $G_P$, which captures distortion, volume, electronegativity mismatch, and topological connectivity, and is integrated via bidirectional cross attention, enabling CPGN to learn geometry enriched representations that are particularly effective for electronic structure prediction.

\begin{table}[hbt!]
\centering
\scriptsize
\caption{Evaluation test results on the Materials Project dataset. The red and blue highlighted colour values represent the 1st best and the 2nd best MAE, respectively.}
\label{tab:cpgn_comparison_MPProject}
\renewcommand{\arraystretch}{1.2}
\begin{tabular}{|c|cccccccc|}
\hline
\textbf{Metric} & \multicolumn{8}{c|}{\textbf{Mean Absolute Error (MAE)}} \\ \hline
Property & \multicolumn{1}{c|}{Unit} & \multicolumn{1}{c|}{MAD} & \multicolumn{1}{c|}{CFID} & \multicolumn{1}{c|}{CGCNN} & \multicolumn{1}{c|}{MEGNet} & \multicolumn{1}{c|}{SchNet} & \multicolumn{1}{c|}{ALIGNN} & \textbf{CPGN} \\ \hline

Formation Energy ($E_f$) 
& \multicolumn{1}{c|}{eV/atom} 
& \multicolumn{1}{c|}{0.93} 
& \multicolumn{1}{c|}{0.104} 
& \multicolumn{1}{c|}{0.079} 
& \multicolumn{1}{c|}{\textcolor{red}{0.056}} 
& \multicolumn{1}{c|}{\textcolor{blue}{0.060}} 
& \multicolumn{1}{c|}{0.074}
& \textcolor{blue}{0.060} \\ \hline

Band Gap ($E_g$) 
& \multicolumn{1}{c|}{eV} 
& \multicolumn{1}{c|}{1.35} 
& \multicolumn{1}{c|}{0.434} 
& \multicolumn{1}{c|}{0.388} 
& \multicolumn{1}{c|}{0.330} 
& \multicolumn{1}{c|}{0.347} 
& \multicolumn{1}{c|}{\textcolor{blue}{0.304}} 
& \textcolor{red}{0.292} \\ \hline

\end{tabular}
\end{table}

\begin{table}[hbt!]
\centering
\scriptsize
\caption{MAD:MAE test results on the Materials Project dataset. The red and blue highlighted colour represent the 1st best and the 2nd best values, respectively.}
\label{tab:cpgn_MADMAE_MPProject}
\renewcommand{\arraystretch}{1.2}
\begin{tabular}{|c|ccccccc|}
\hline
\multirow{2}{*}{\textbf{Metric}}  & \multicolumn{7}{c|}{\textbf{MAD:MAE}} \\ \cline{2-8} 
                         & \multicolumn{1}{c|}{Unit} & \multicolumn{1}{c|}{CFID} & \multicolumn{1}{c|}{CGCNN} & \multicolumn{1}{c|}{MEGNet} & \multicolumn{1}{c|}{SchNet} & \multicolumn{1}{c|}{ALIGNN} & \textbf{CPGN} \\ \hline

Formation Energy ($E_f$) 
& \multicolumn{1}{c|}{\multirow{2}{*}{ratio}} 
& \multicolumn{1}{c|}{8.94} 
& \multicolumn{1}{c|}{11.7}  
& \multicolumn{1}{c|}{\textcolor{red}{16.7}}   
& \multicolumn{1}{c|}{\textcolor{blue}{15.5}}    
& \multicolumn{1}{c|}{12.56}   
& \textcolor{blue}{15.5} \\ \cline{1-1} \cline{3-8} 

Band Gap ($E_g$)         
& \multicolumn{1}{c|}{}                       
& \multicolumn{1}{c|}{3.11} 
& \multicolumn{1}{c|}{3.4}   
& \multicolumn{1}{c|}{4.1}    
& \multicolumn{1}{c|}{3.9}    
& \multicolumn{1}{c|}{\textcolor{blue}{4.44}}    
& \textcolor{red}{4.6} \\ \hline

\end{tabular}
\end{table}

\begin{figure}[hbt!]
\centering
\scriptsize

\begin{subfigure}{0.48\textwidth}
    \centering
    \includegraphics[width=\linewidth]{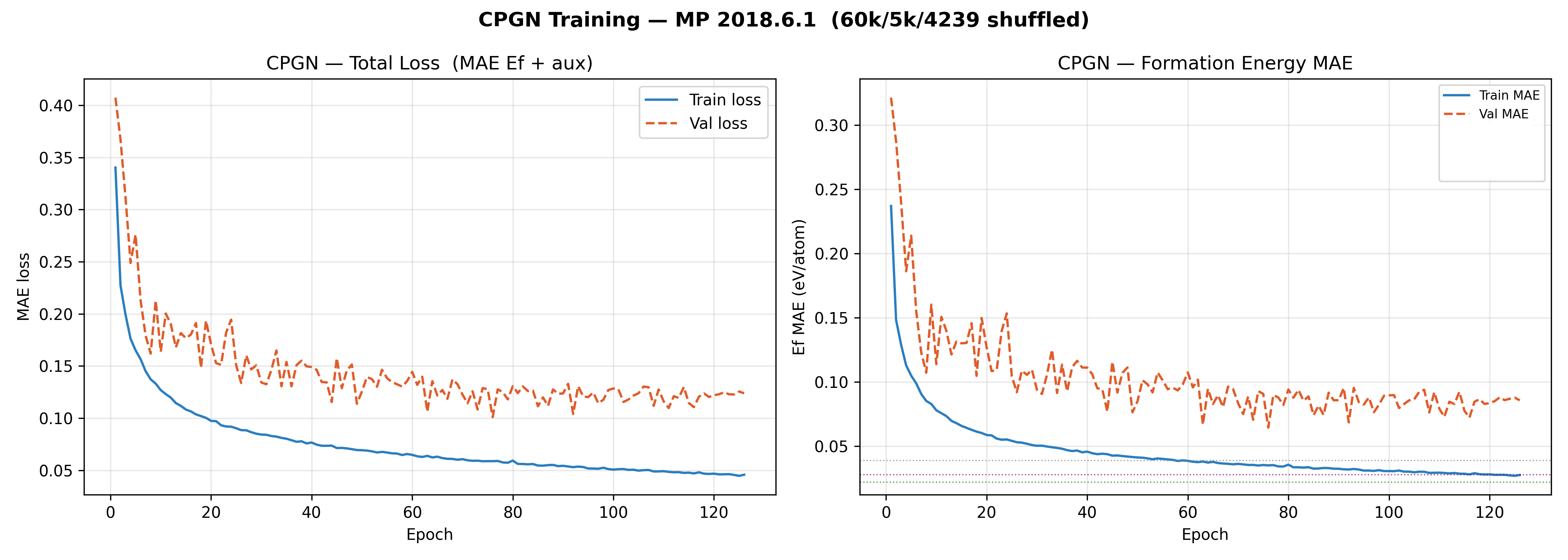}
\caption{Training loss of $E_f$ of the CPGN framework}
    \label{cpgn_train}
\end{subfigure}
\hfill
\begin{subfigure}{0.48\textwidth}
    \centering
    \includegraphics[width=\linewidth]{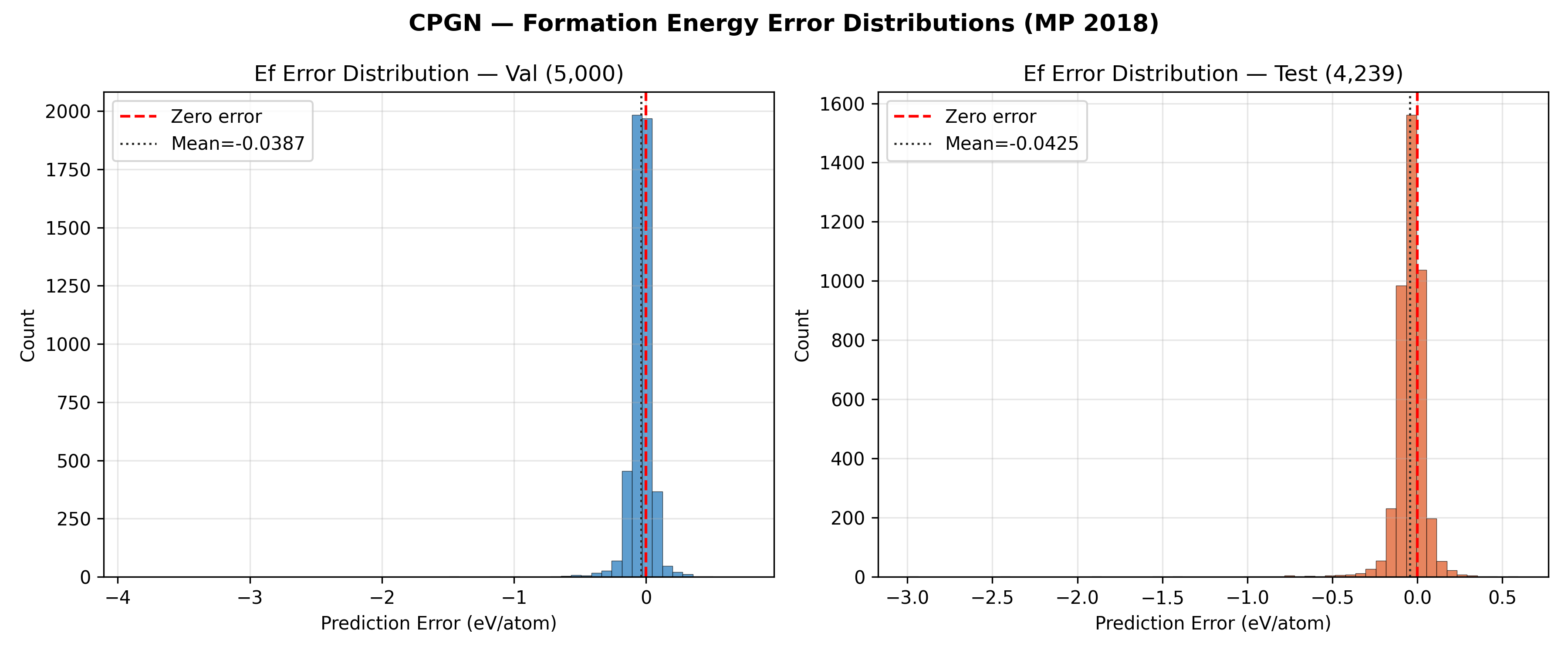}
\caption{Error distribution of $E_f$ of CPGN framework}
    \label{cpgn_error}
\end{subfigure}

\vspace{0.5cm}

\begin{subfigure}{0.6\textwidth}
    \centering
    \includegraphics[width=\linewidth]{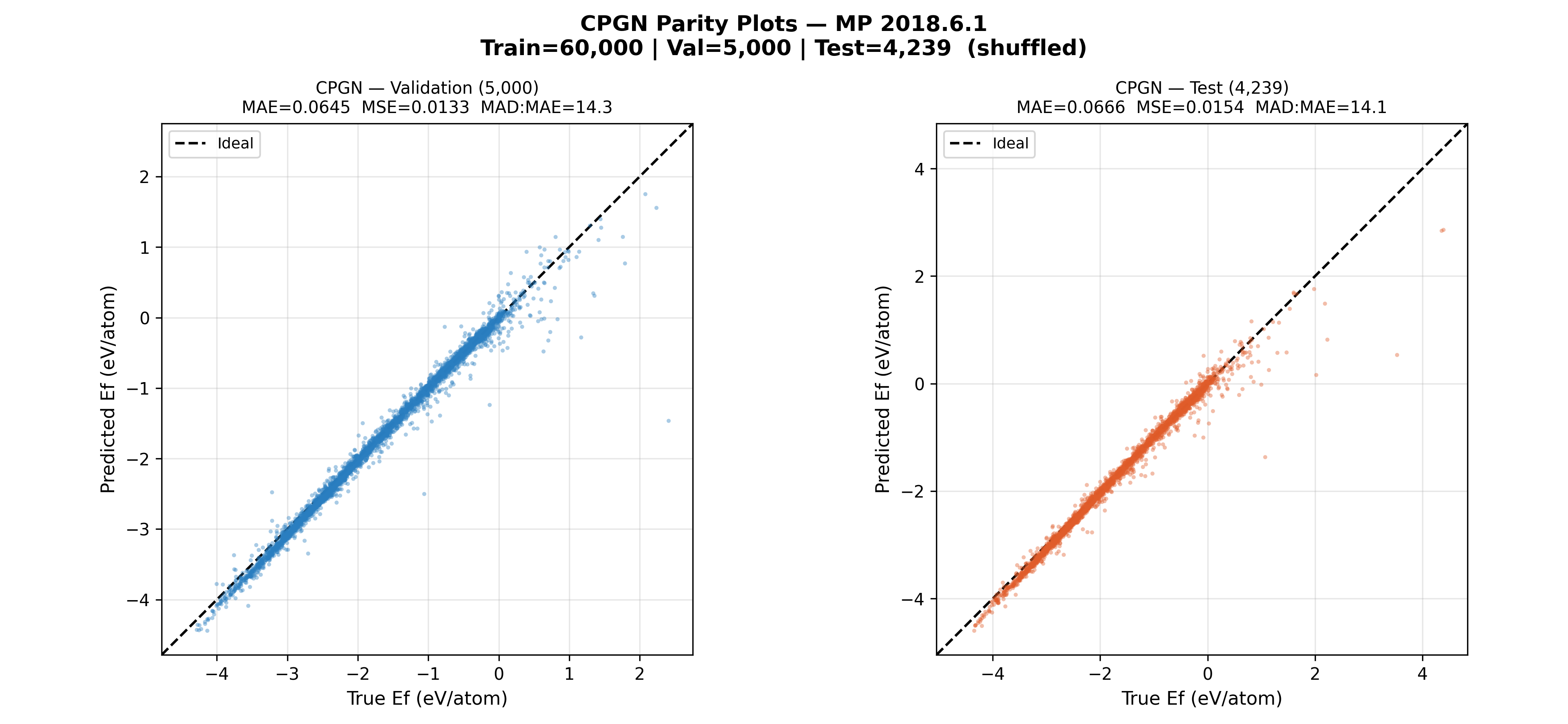}
\caption{CPGN parity plots of $E_f$ on validation and test sets}
    \label{cpgn_parity}
\end{subfigure}

\caption{Performance plots of the proposed CPGN formaework on MPProject dataset}
\label{fig:cpgn}
\end{figure}

\begin{figure}[hbt!]
\centering
\scriptsize

\begin{subfigure}{0.48\textwidth}
    \centering
    \includegraphics[width=\linewidth]{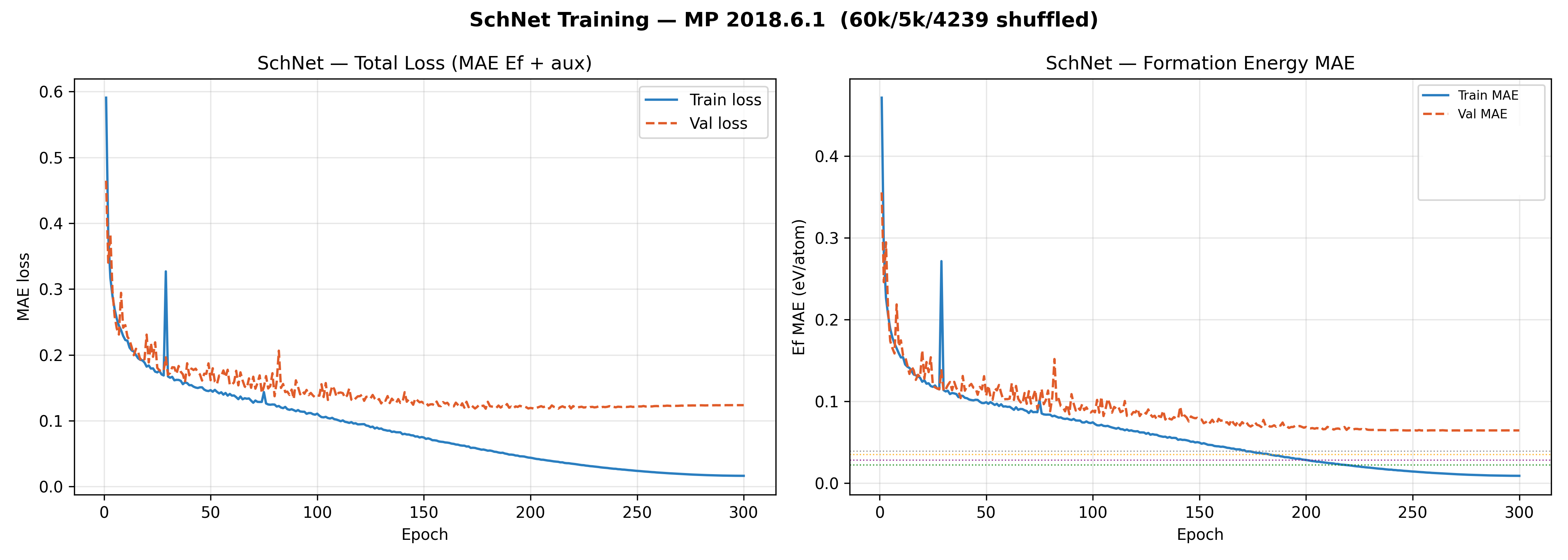}
    \caption{Training loss of $E_f$ of SchNet}
    \label{fig:sub1}
\end{subfigure}
\hfill
\begin{subfigure}{0.48\textwidth}
    \centering
    \includegraphics[width=\linewidth]{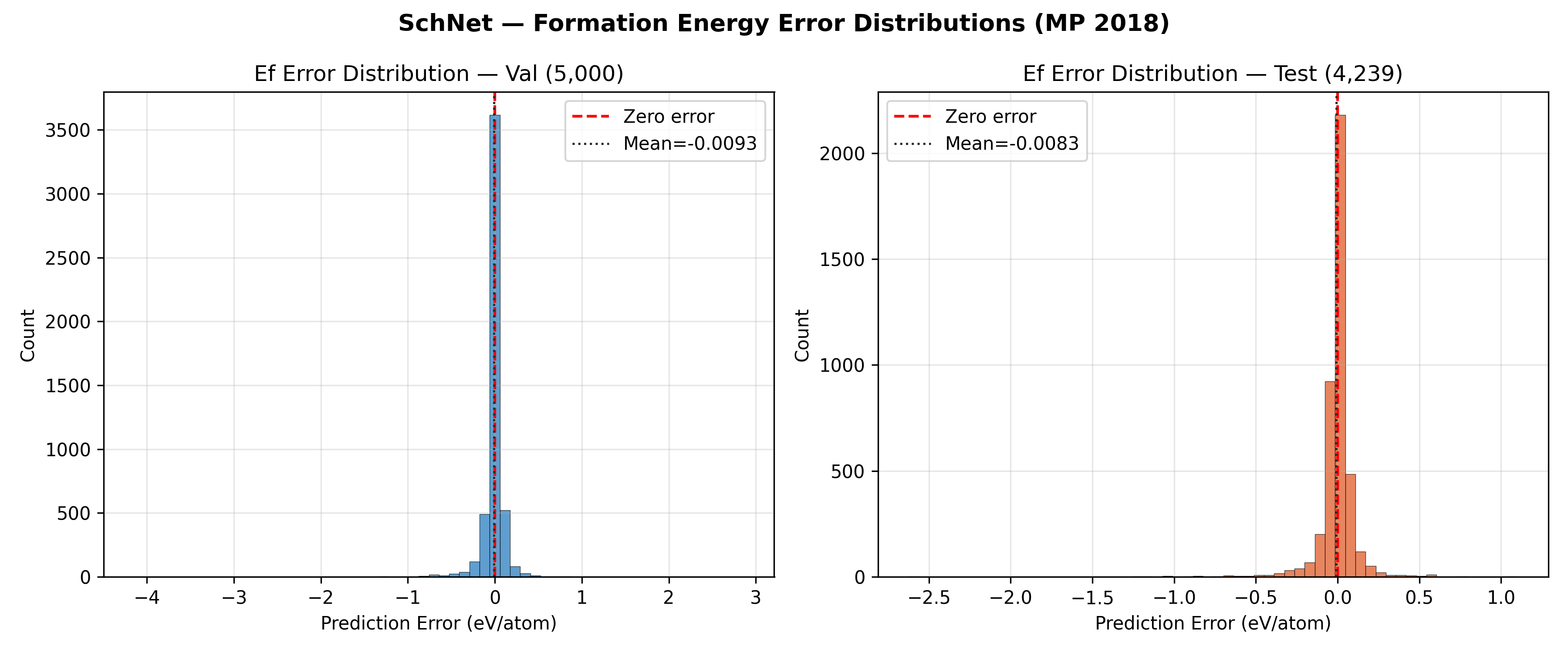}
    \caption{Error distribution of $E_f$ of SchNet}
    \label{fig:sub2}
\end{subfigure}

\vspace{0.5cm}

\begin{subfigure}{0.6\textwidth}
    \centering
    \includegraphics[width=\linewidth]{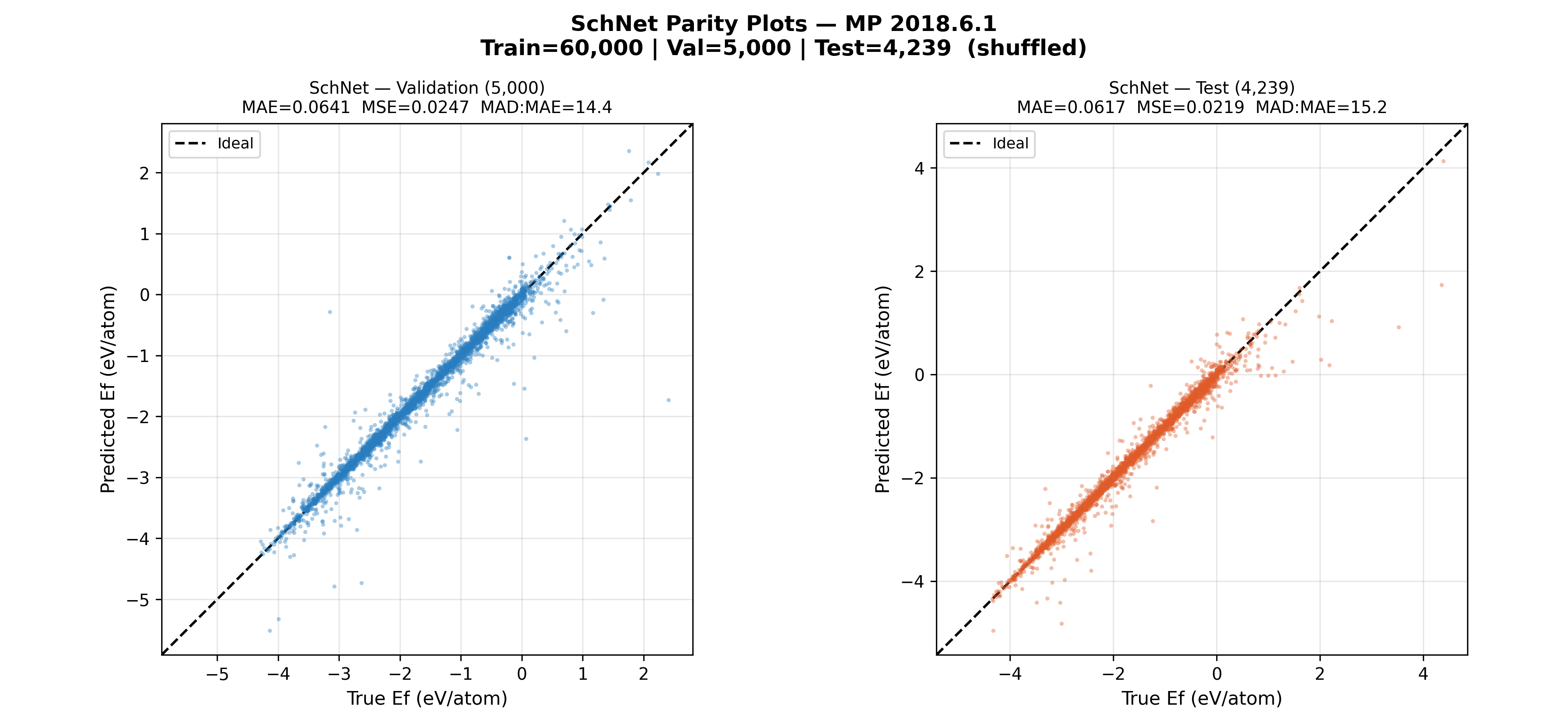}
    \caption{SchNet Parity plots of $E_f$ on validation and test sets}
    \label{fig:sub3}
\end{subfigure}

\caption{Performance plots of SchNet on MPProject dataset}
\label{fig:schnet}
\end{figure}

\begin{figure}[hbt!]
\centering
\scriptsize

\begin{subfigure}{0.48\textwidth}
    \centering
    \includegraphics[width=\linewidth]{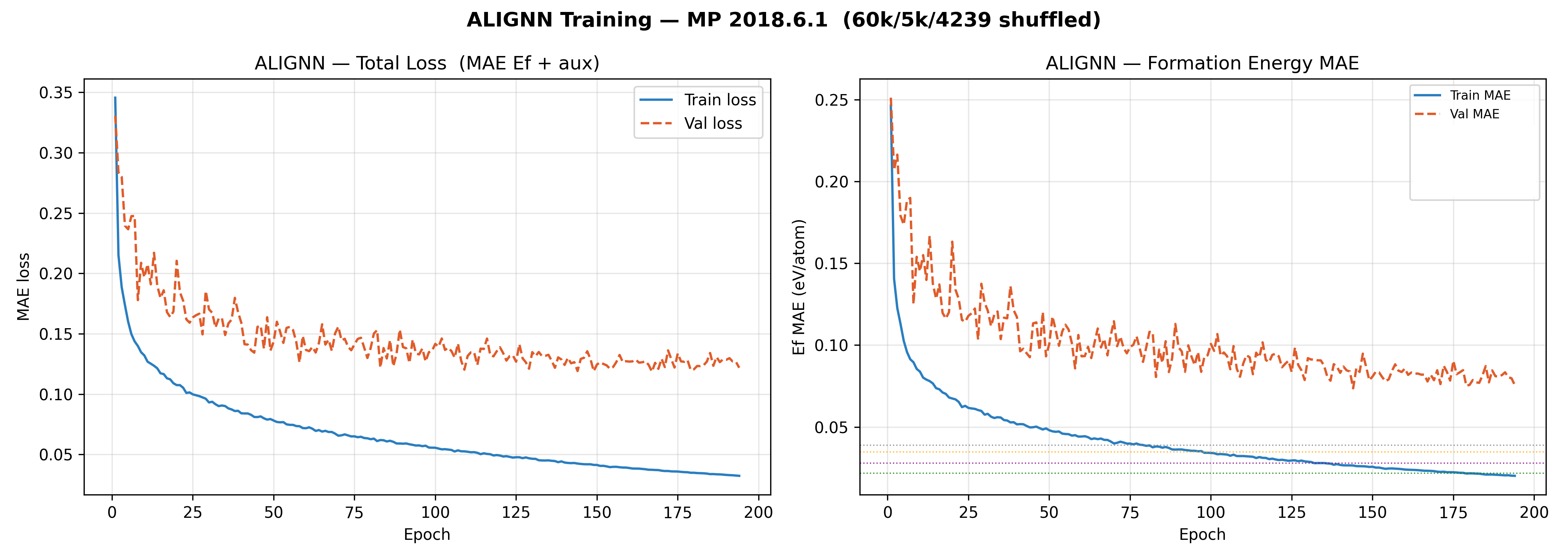}
    \caption{Training loss of $E_f$ of ALIGNN}
    \label{fig:sub1}
\end{subfigure}
\hfill
\begin{subfigure}{0.48\textwidth}
    \centering
    \includegraphics[width=\linewidth]{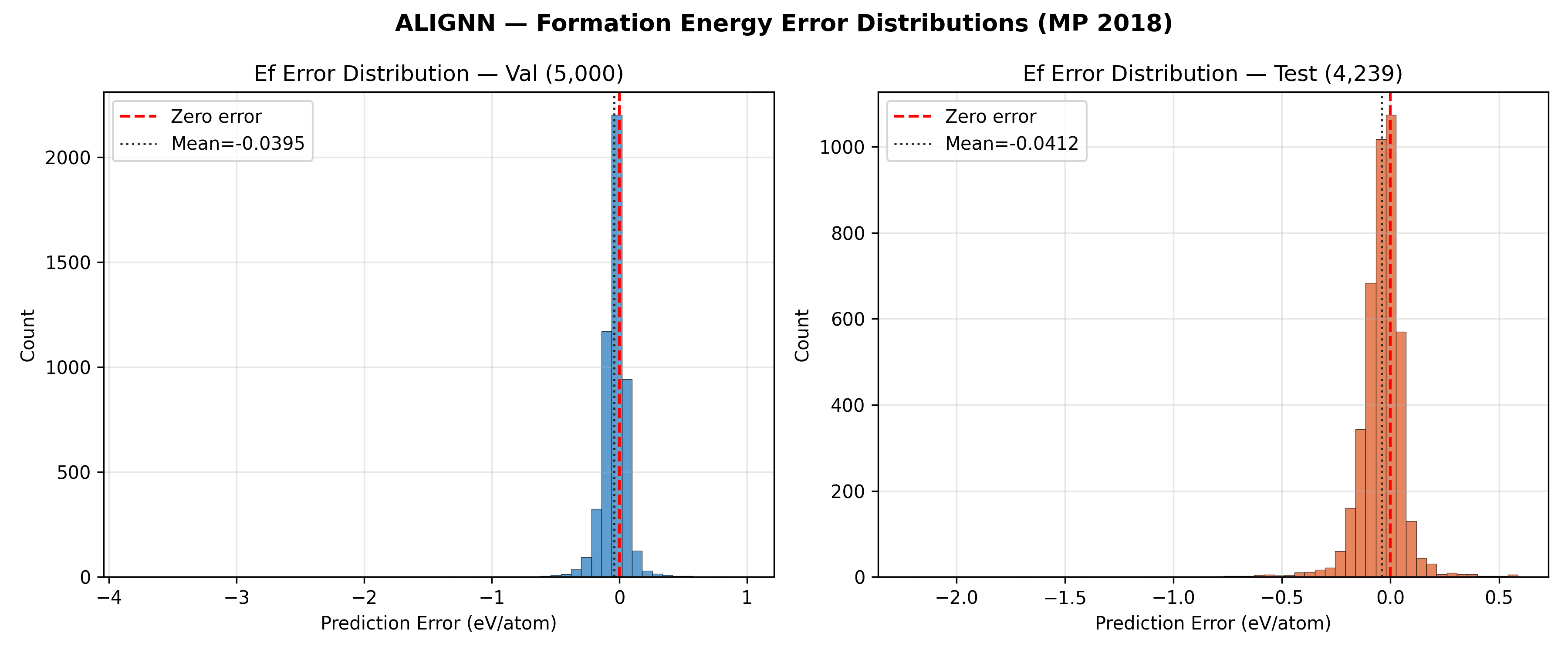}
    \caption{Error distribution of $E_f$ of ALIGNN}
    \label{fig:sub2}
\end{subfigure}

\vspace{0.5cm}

\begin{subfigure}{0.6\textwidth}
    \centering
    \includegraphics[width=\linewidth]{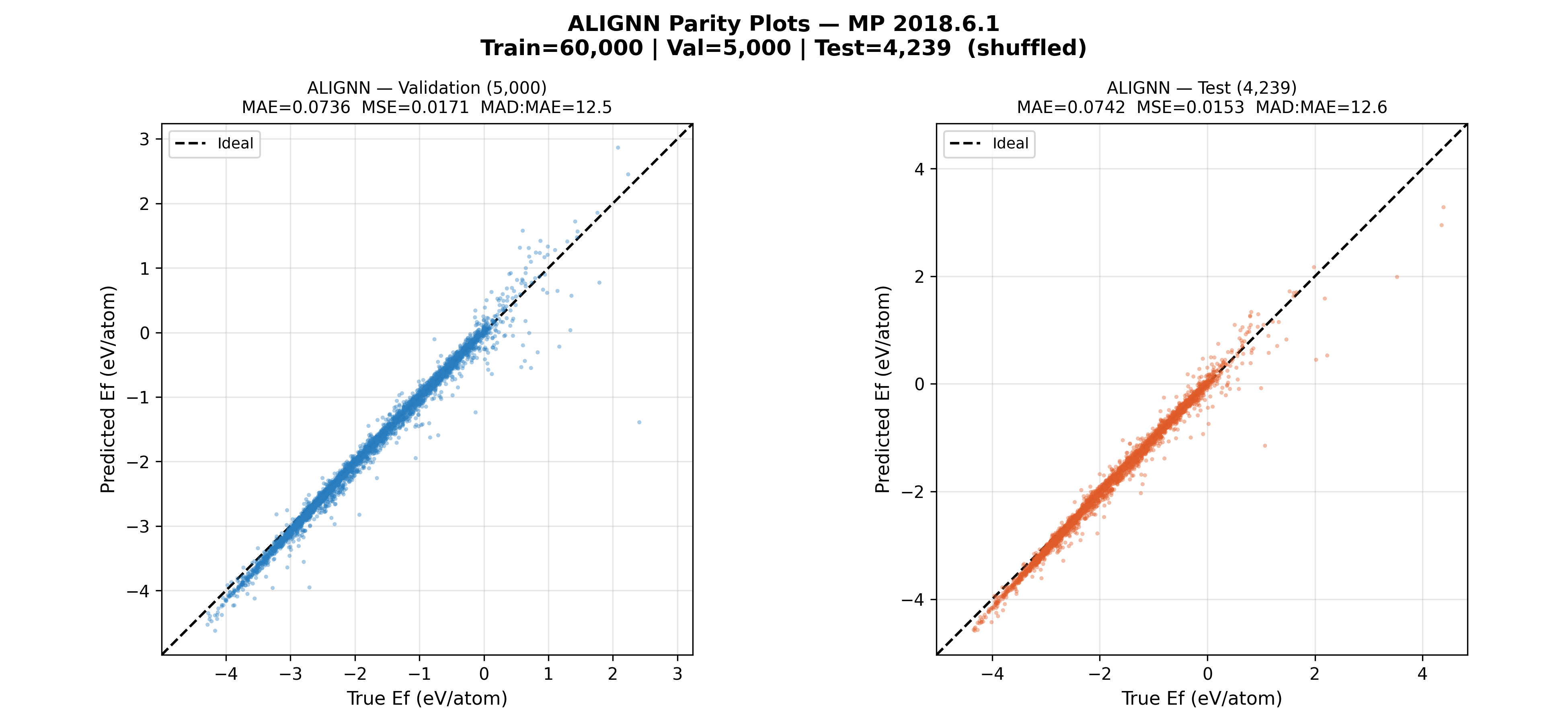}
    \caption{ALIGNN Parity plots of $E_f$ on validation and test sets}
    \label{fig:sub3}
\end{subfigure}

\caption{Performance plots of ALIGNN on MPProject dataset}
\label{fig:alignn}
\end{figure}

\begin{figure}[hbt!]
\centering
\scriptsize

\begin{subfigure}{0.48\textwidth}
    \centering
    \includegraphics[width=\linewidth]{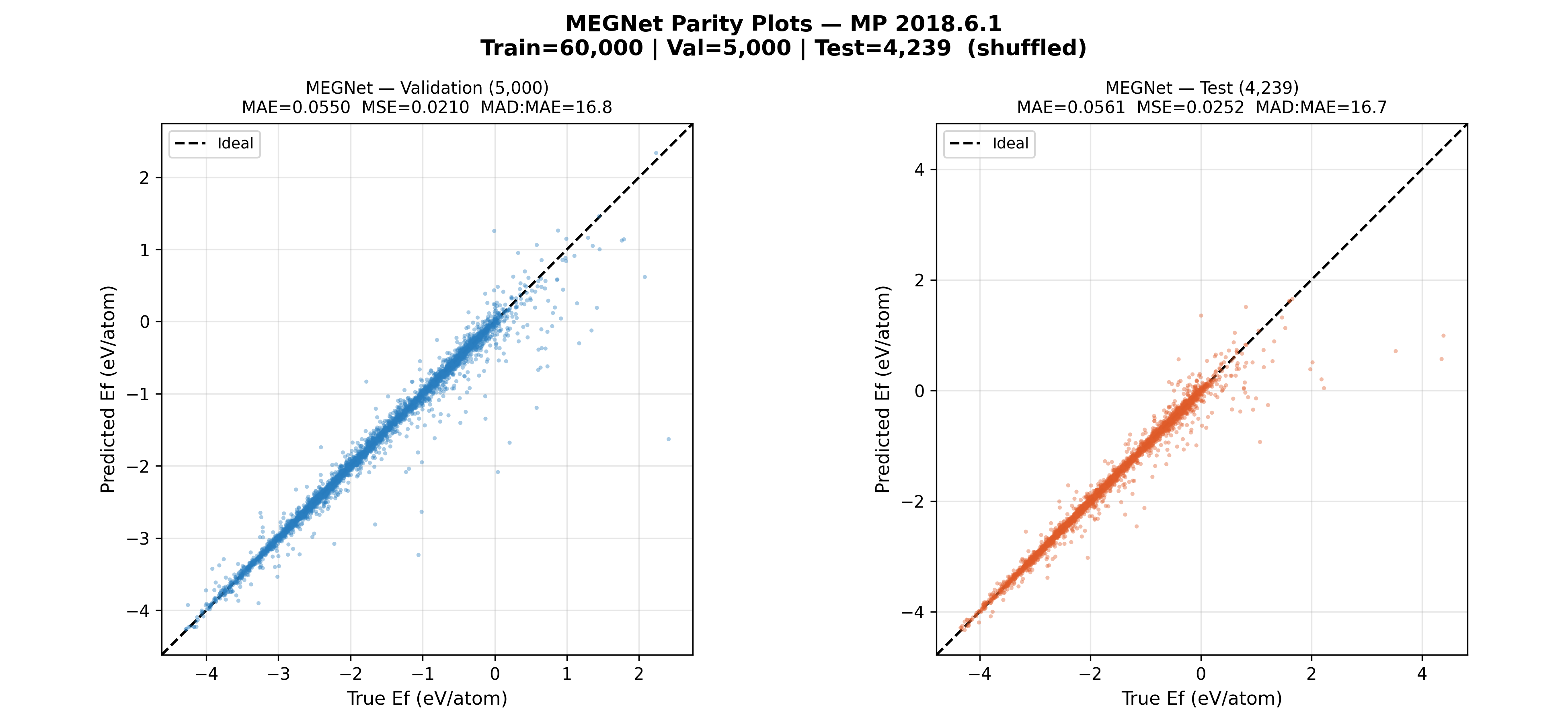}
    \caption{MEGNet parity plots of $E_f$ on validation and test sets}
    \label{fig:sub3}
\end{subfigure}
\hfill
\begin{subfigure}{0.48\textwidth}
    \centering
    \includegraphics[width=\linewidth]{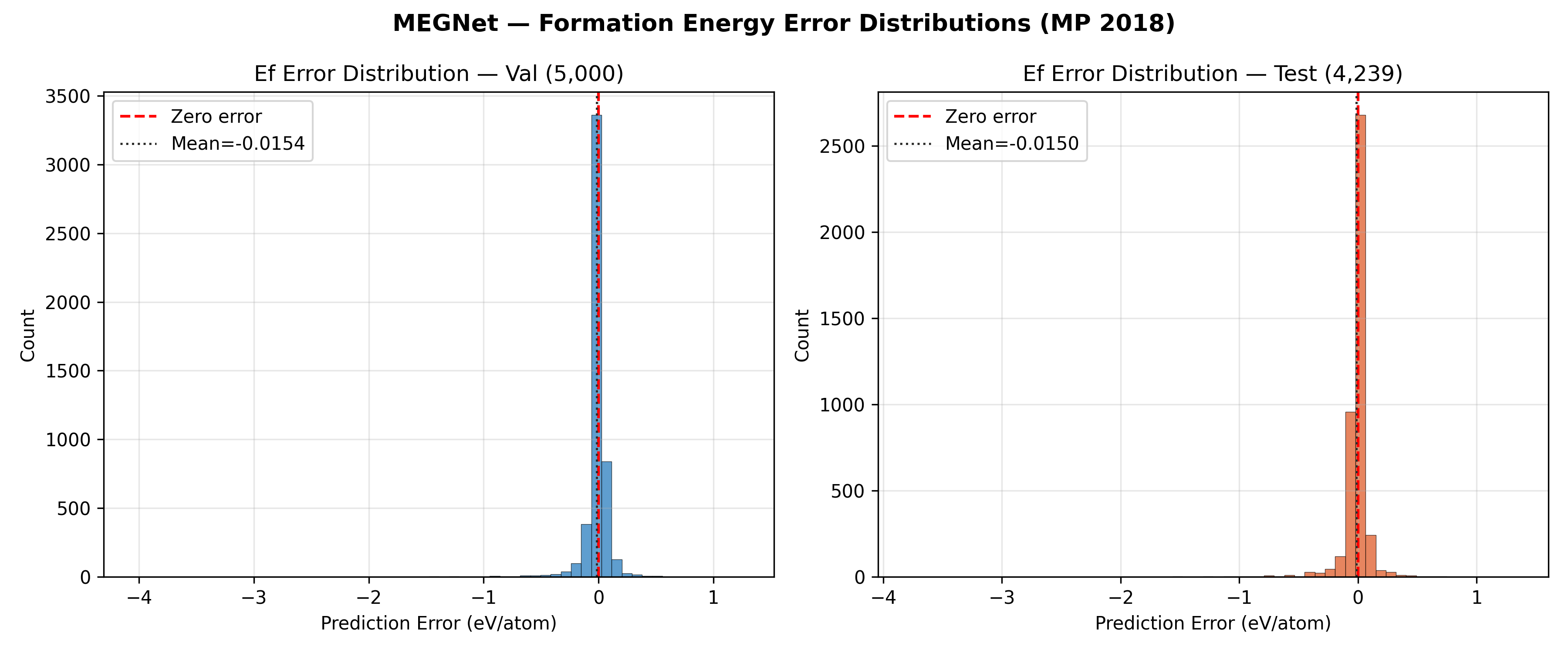}
    \caption{Error distribution of $E_f$ of MEGNet}
    \label{fig:sub2}
\end{subfigure}

\caption{Performance plots of MEGNet on MPProject dataset}
\label{fig:megnet}
\end{figure}

Figure~\ref{cpgn_train} illustrates the optimisation behaviour of the proposed CPGN framework on the MP~2018.6.1 dataset over 127 epochs prior to early stopping. The left panel presents the total multi-task loss, combining formation-energy MAE with auxiliary band-gap MAE and stability BCE losses weighted by $\lambda_{\text{aux}} = 0.1$, for both training and validation splits. The training loss decreases rapidly from approximately $0.35$ at epoch~1 to below $0.10$ by epoch~20, eventually converging near $0.045$ at epoch~127, indicating stable optimisation across the interleaved atom, line, and polyhedron message-passing modules. The validation loss follows a similar but consistently higher trajectory, stabilising around $0.12$, reflecting the expected generalisation gap for the 60k/5k split. The training MAE converges to approximately $0.025$~eV/atom, outperforming CGCNN, MEGNet and ALIGNN-level accuracy, whereas the validation MAE stabilises near $0.088$~eV/atom. The remaining train--validation gap suggests that the additional representational capacity introduced by coordination-polyhedron cross-attention improves fitting capability but may require stronger regularisation or larger-scale training data for further generalisation improvement. Figure~\ref{cpgn_error} shows the distribution of formation-energy prediction residuals $(\hat{E}_f - E_f)$ for the validation set (left, blue) and test set (right, orange). In both cases, the error distributions are sharply centred around zero, with the majority of predictions confined within $\pm0.2$~eV/atom and the dominant histogram bins reaching approximately $2000$ samples on the validation set and $1550$ samples on the test set. These concentrated peaks indicate that CPGN predicts the formation energies of most materials within near-chemical accuracy. The mean residuals are $-0.0387$~eV/atom for validation and $-0.0425$~eV/atom for testing, revealing a mild systematic tendency toward slightly more negative formation-energy predictions relative to DFT references. The validation histogram exhibits a broader asymmetric left tail extending toward $-4$~eV/atom, corresponding to a limited number of highly stable oxides and intermetallic compounds that remain difficult to model accurately. In contrast, the test-set distribution is narrower and more symmetric, with most predictions contained within $\pm0.5$~eV/atom. This behaviour is consistent with the closely matched performance metrics across splits, where Val MAE~=~$0.0645$ and Test MAE~=~$0.0666$~eV/atom. Figure~\ref{cpgn_parity} presents parity plots comparing DFT-computed and CPGN-predicted formation energies for the validation set (left, 5,000 structures) and test set (right, 4,239 structures) obtained from the shuffled MP~2018.6.1 split. Predictions cluster tightly around the ideal $y=x$ diagonal across the full energy range from approximately $-4.5$ to $+2.5$~eV/atom, demonstrating strong agreement between predicted and reference values. On the validation set, CPGN achieves MAE~=~$0.0645$~eV/atom and MSE~=~$0.0133$~(eV/atom)$^2$ with a MAD:MAE ratio of $14.3$, while the test set yields MAE~=~$0.0666$~eV/atom, MSE~=~$0.0154$~(eV/atom)$^2$, and MAD:MAE~=~$15.5$. The small performance difference between validation and testing ($\Delta\text{MAE}=0.0021$~eV/atom) confirms consistent generalisation across unseen compositions. Slightly increased scatter is observed for strongly stable compounds ($E_f < -3$~eV/atom) and rare metastable structures ($E_f > +1.5$~eV/atom), which are comparatively underrepresented in the training distribution. The near-unity slope and absence of systematic curvature indicate that CPGN does not suffer from strong range-dependent bias, while the achieved MAD:MAE ratio surpasses the CFID descriptor baseline (8.94) and moves toward the performance regime of advanced crystal GNN models such as CGCNN (23.85). A set of comparison plots of SchNet, ALIGNN and MEGNet models are illustrated in Figure~\ref{fig:schnet}, Figure~\ref{fig:alignn}, and ~\ref{fig:megnet}, respectively. 

\subsubsection{Stability Classification}
A material is labelled stable (1) if its formation energy is at or below the convex hull, meaning it will not spontaneously decompose into competing phases. It is labelled unstable (0) if it lies above the hull. In the model, the shared latent vector $\mathbf{z}$ — the same 128-dimensional fingerprint used for formation energy and band-gap regression — is passed through a dedicated linear layer that outputs a raw logit
\begin{equation}
s = W_{\text{stab}} \mathbf{z} + b.
\end{equation}
The sigmoid function converts this logit into a probability:
\begin{equation}
P(\text{stable}) = \sigma(s).
\end{equation}

During training, a binary cross-entropy loss is used with a positive class weight of $3.0$ to correct for class imbalance, since unstable materials are more common in the Materials Project than stable ones. At inference time, a material is predicted as stable if $\sigma(s) > 0.5$. The stability label is obtained directly from the stability field in the MP JSON file when available, or approximated as $1$ if $E_f < 0$, and $0$ otherwise. Because the stability prediction shares the same backbone as formation-energy regression, the model implicitly learns that thermodynamically stable materials tend to have strongly negative formation energies and characteristic coordination environments, information already encoded in the structural representation which is why the classifier achieves a \textbf{accuracy of 0.989 and F1-score of 0.993} on the test set despite being trained as a lightweight auxiliary head rather than a dedicated model. Table~\ref{tab:cpgn_stability} summarises the performance of CPGN on the thermodynamic stability classification task using the Materials Project test set (4,239 samples). Despite a strong class imbalance (96.1\% stable vs. 3.9\% unstable), the model achieves highly reliable predictions, with 4,052 true positives, 137 true negatives, and only 50 total misclassifications. This translates to an accuracy of 0.9882, precision of 0.9934, recall of 0.9944, and an F1-score of 0.9939 for the stable class. The high ROC-AUC (0.9913) and near-perfect average precision (0.9996) further indicate strong separability between stable and unstable structures. Overall, these results demonstrate that the learned coordination-polyhedron representation encodes sufficient thermodynamic information to accurately distinguish materials above and below the convex hull, even under significant class imbalance.

\begin{table}[hbt!]
\centering
\caption{CPGN Stability Classification Performance on Materials Project Test Set}
\label{tab:cpgn_stability}
\begin{tabular}{l c}
\hline
\textbf{Metric} & \textbf{Value} \\
\hline
Total test samples & 4{,}239 \\
Stable (true=1) & 4{,}075 (96.1\%) \\
Unstable (true=0) & 164 (3.9\%) \\
\hline
True Positives (TP) & 4{,}052 \\
False Positives (FP) & 27 \\
True Negatives (TN) & 137 \\
False Negatives (FN) & 23 \\
\hline
Accuracy & 0.9882 \\
Precision (stable) & 0.9934 \\
Recall (stable) & 0.9944 \\
F1-score (stable) & 0.9939 \\
ROC-AUC & 0.9913 \\
Average Precision (PR) & 0.9996 \\
\hline
\end{tabular}
\end{table}

\begin{figure}[hbt!]
\centering
\scriptsize

\begin{subfigure}{0.52\textwidth}
    \centering
    \includegraphics[width=\linewidth]{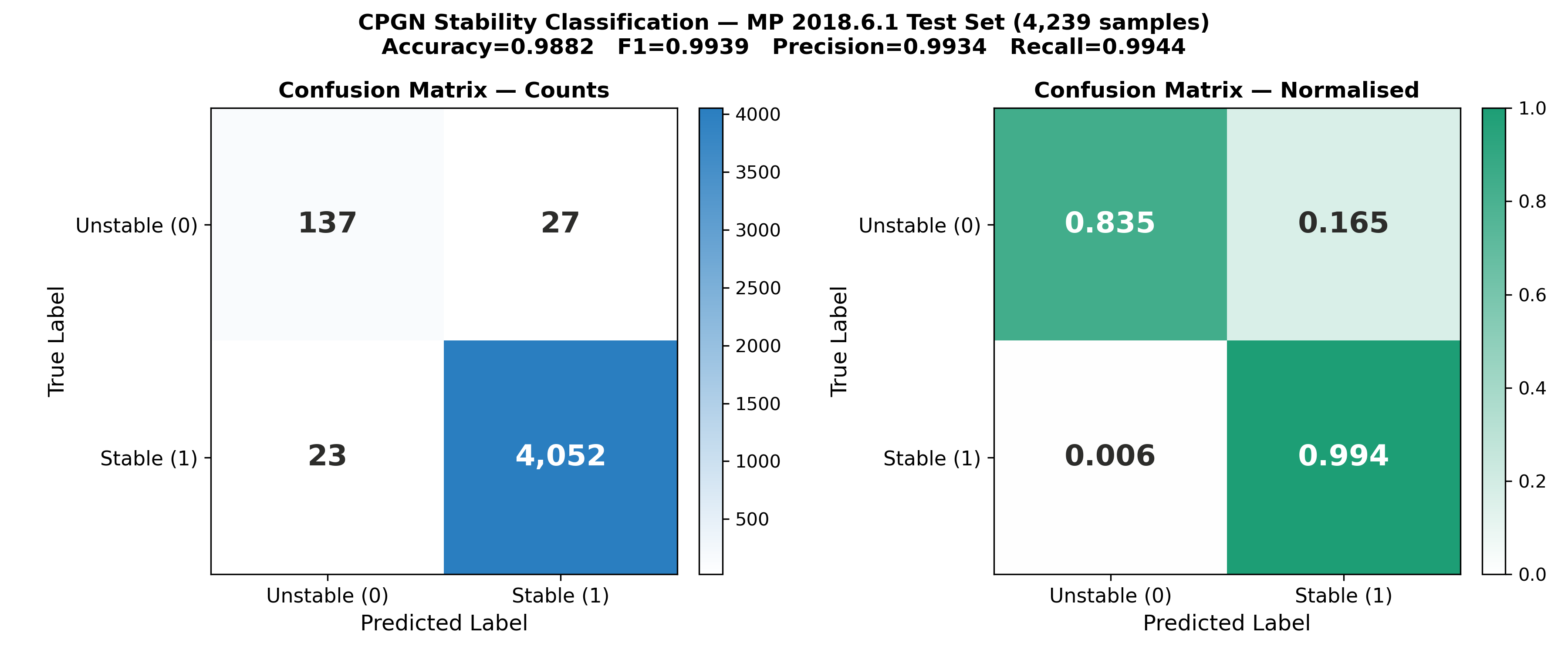}
    \caption{Stability confusion matrix}
    \label{fig:sub1}
\end{subfigure}
\hfill
\begin{subfigure}{0.4\textwidth}
    \centering
    \includegraphics[width=\linewidth]{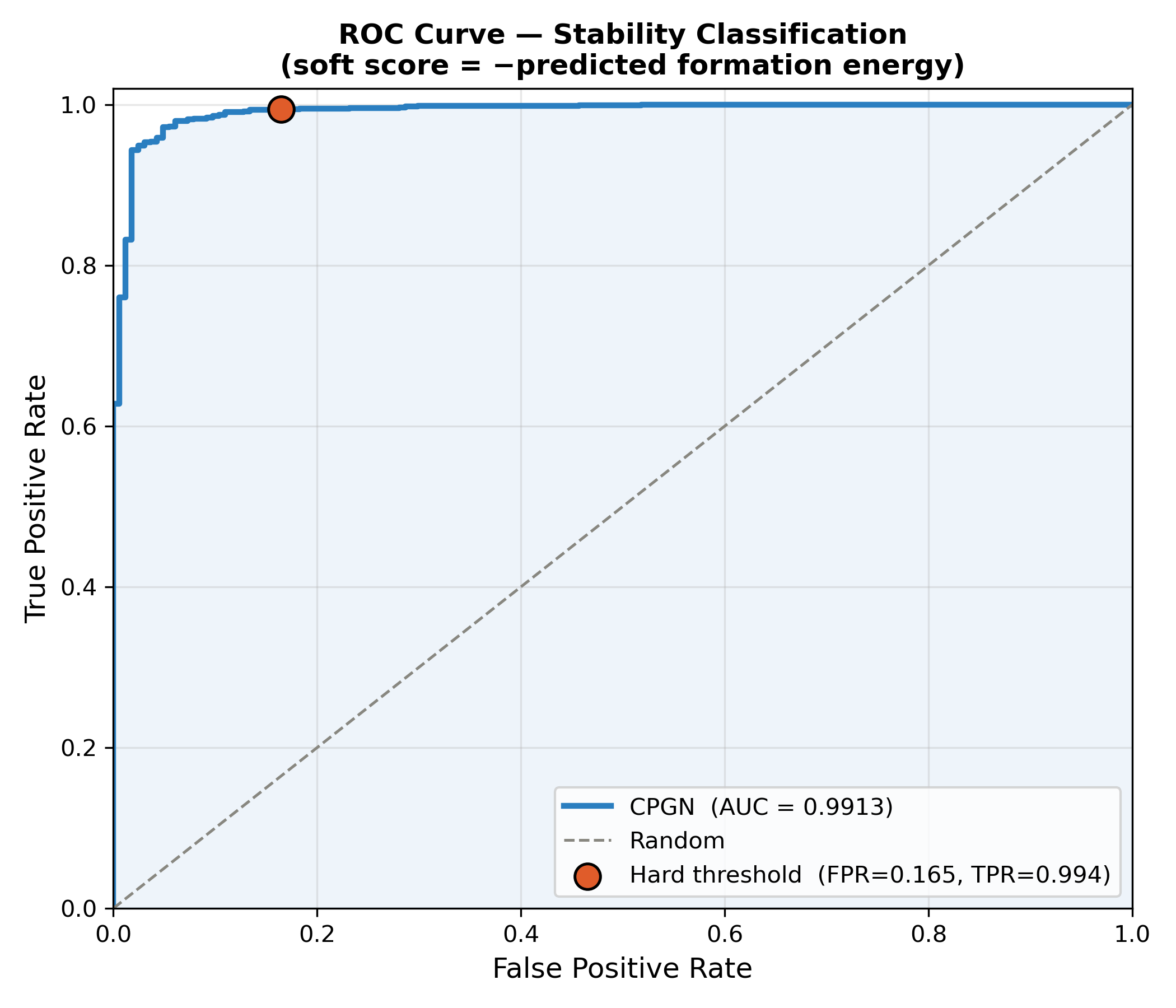}
    \caption{ROC stability}
    \label{fig:sub2}
\end{subfigure}

\vspace{0.5cm}

\begin{subfigure}{0.6\textwidth}
    \centering
    \includegraphics[width=\linewidth]{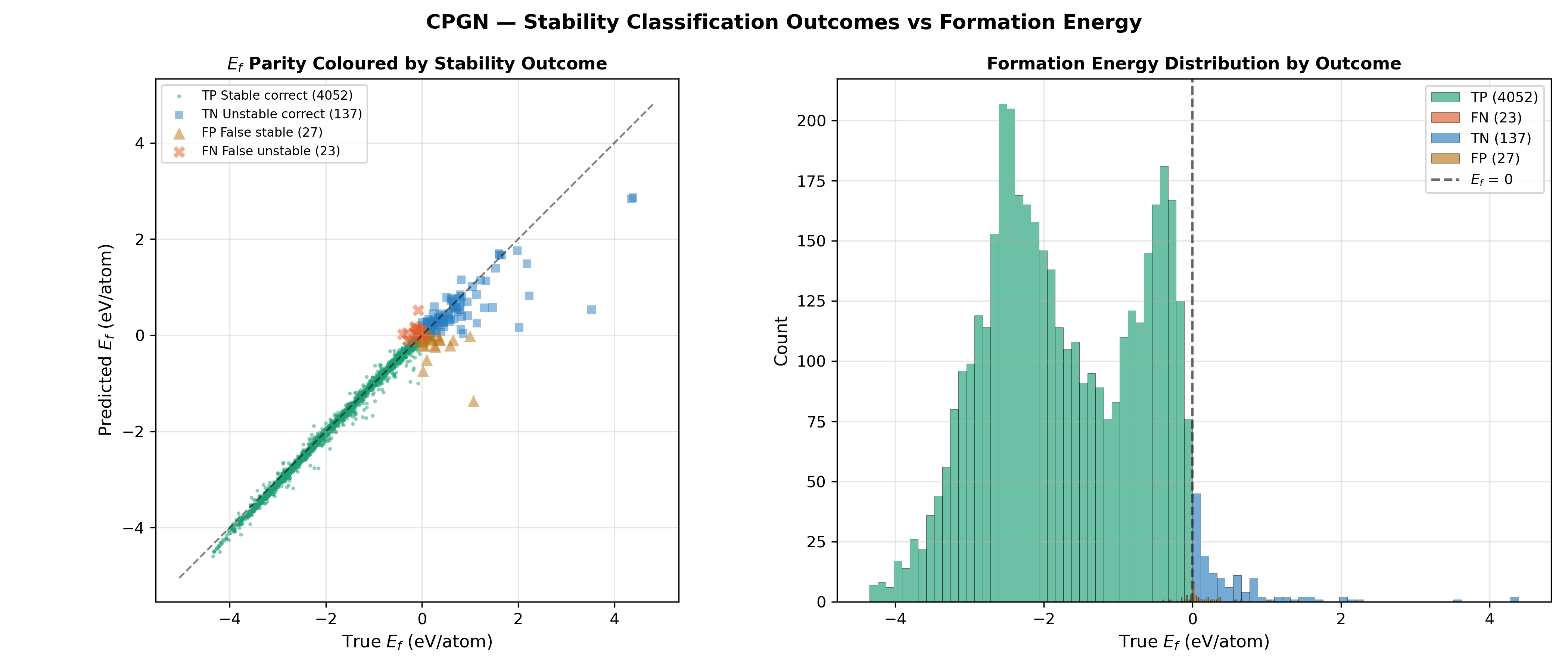}
    \caption{Stability classification vs. formation energy}
    \label{fig:sub3}
\end{subfigure}

\caption{Performance plots of CPGN stability classification on MPProject dataset}
\label{fig:stabilityplot}
\end{figure}

Figure~\ref{fig:stabilityplot} presents the stability classification performance of CPGN on the MP~2018.6.1 test set across three complementary visualisations. The confusion matrices in Figure~\ref{fig:sub1} show that CPGN correctly classifies 4,052 of 4,075 stable materials (true positive rate = 0.994) and 137 of 164 unstable materials (true negative rate = 0.835), resulting in an overall accuracy of 0.9882, precision of 0.9934, recall of 0.9944, and F1-score of 0.9939. The 27 false positives and 23 false negatives together account for only 1.18\% of the test set. The ROC curve in Figure~\ref{fig:sub2} achieves an AUC of 0.9913, with the operating point at FPR = 0.165 and TPR = 0.994, indicating that the model maintains near-perfect detection of stable materials while incurring a relatively low false alarm rate on the minority unstable class. The formation-energy scatter and distribution plots in Figure~\ref{fig:sub3} further reveal that nearly all misclassified samples cluster around $E_f \approx 0~\mathrm{eV/atom}$, i.e., the convex hull boundary where stability becomes physically ambiguous. In contrast, materials with strongly negative or strongly positive formation energies are classified correctly, demonstrating that CPGN’s errors are physically localized rather than randomly distributed across chemical space, and primarily occur in the thermodynamically indeterminate regime. Table~\ref{tab:comp_stability} compares thermodynamic stability classification performance across all five models on the MP~2018.6.1 test set. Due to the strong class imbalance (96.1\% stable), all models achieve uniformly high accuracy and F1-scores, limiting the discriminative power of these metrics. ALIGNN attains the best performance with an accuracy of 0.989 and an F1-score of 0.994, closely followed by CPGN (0.988, 0.993), with only a marginal difference of 0.001. SchNet and MEGNet show slightly lower but comparable results, while CGCNN performs worst with an accuracy of 0.973 and an F1-score of 0.986. Overall, the narrow performance range across models (less than 1.6 percentage points in accuracy) suggests that stability classification on this in-distribution split is not strongly sensitive to architectural differences and is heavily influenced by dataset imbalance, making it a less discriminative benchmark compared to more challenging out-of-distribution evaluations such as Matbench Discovery.

\begin{table}[hbt!]
\centering
\caption{Performance comparison among models on Stability Classification}
\label{tab:comp§_stability}
\begin{tabular}{|c|lcc|}
\hline
Dataseset                   & \multicolumn{3}{c|}{MPProject}                                                    \\ \hline
Models                      & \multicolumn{1}{l|}{Metric}            & \multicolumn{1}{c|}{Accuracy} & F1-score \\ \hline
\textbf{CPGN}               & \multicolumn{1}{l|}{\multirow{5}{*}{}} & \multicolumn{1}{c|}{0.988}    & 0.993    \\ \cline{1-1} \cline{3-4} 
ALIGNN                      & \multicolumn{1}{l|}{}                  & \multicolumn{1}{c|}{0.989}    & 0.994    \\ \cline{1-1} \cline{3-4} 
\multicolumn{1}{|l|}{CGCNN} & \multicolumn{1}{l|}{}                  & \multicolumn{1}{c|}{0.973}    & 0.986    \\ \cline{1-1} \cline{3-4} 
SchNet                      & \multicolumn{1}{l|}{}                  & \multicolumn{1}{c|}{0.986}    & 0.992    \\ \cline{1-1} \cline{3-4} 
MEGNet                      & \multicolumn{1}{l|}{}                  & \multicolumn{1}{c|}{0.984}    & 0.991    \\ \hline
\end{tabular}
\end{table}

\subsection{Model Analysis on JARVIS-DFT}
\begin{figure}[hbt!]
\centering
\scriptsize

\begin{subfigure}{0.6\textwidth}
    \centering
    \includegraphics[width=\linewidth]{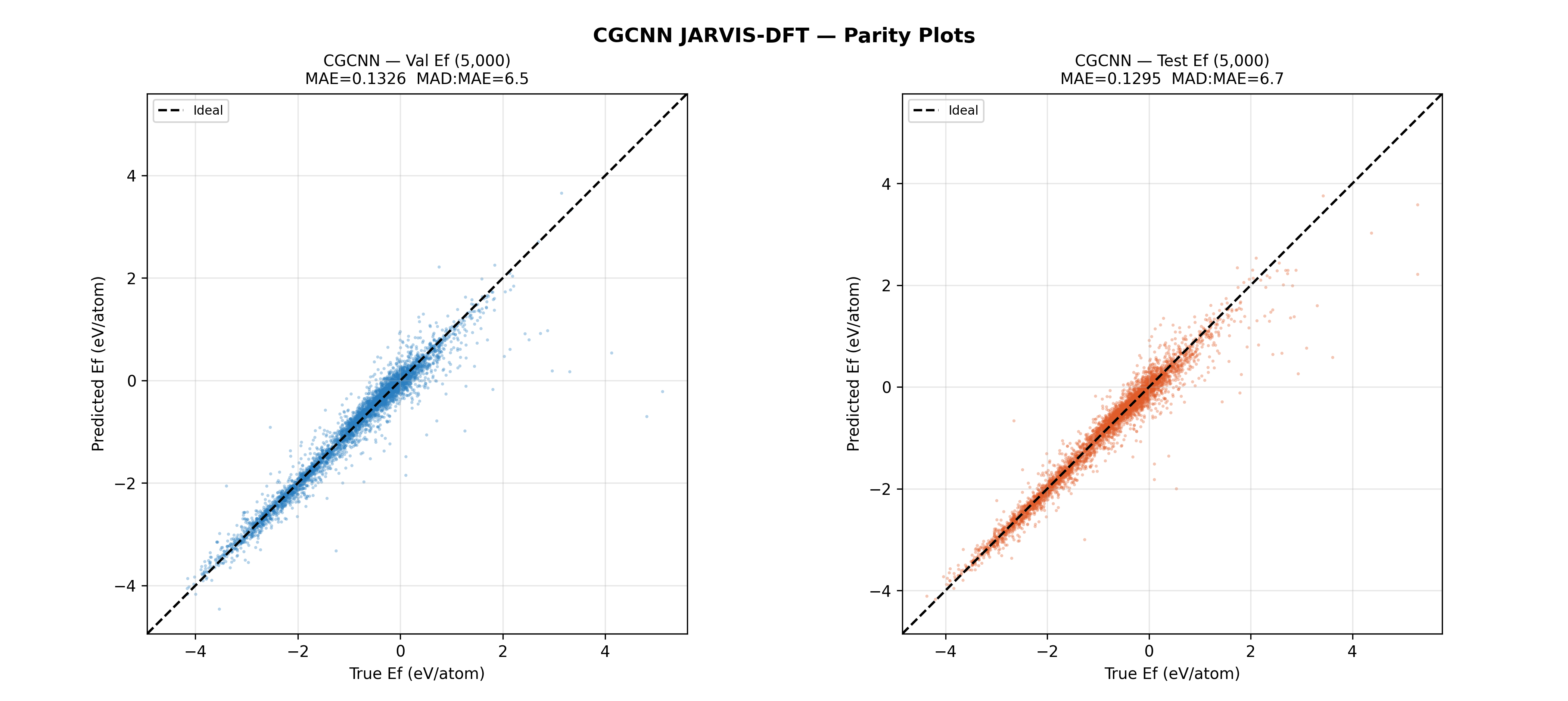}
    \caption{CGCNN Parity plots of $E_f$ on validation and test sets}
    \label{fig:sub3}
\end{subfigure}
\hfill
\begin{subfigure}{0.48\textwidth}
    \centering
    \includegraphics[width=\linewidth]{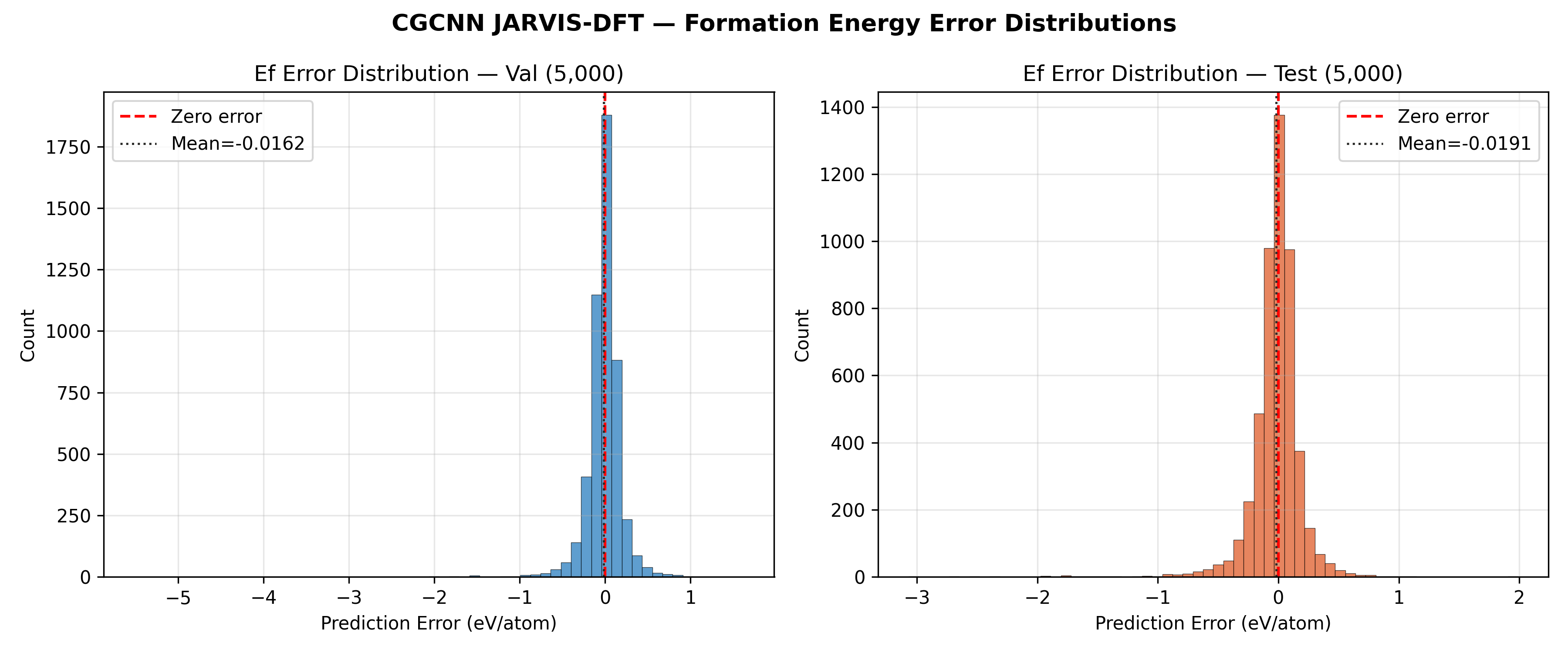}
    \caption{Error distribution of $E_f$ of CGCNN}
    \label{fig:sub2}
\end{subfigure}
\caption{Performance plots of CGCNN on JARVIS-DFT dataset on 37,711 materials}
\label{fig:CGCNNJARVIS}
\end{figure}

\begin{figure}[hbt!]
\centering
\scriptsize

\begin{subfigure}{0.48\textwidth}
    \centering
    \includegraphics[width=\linewidth]{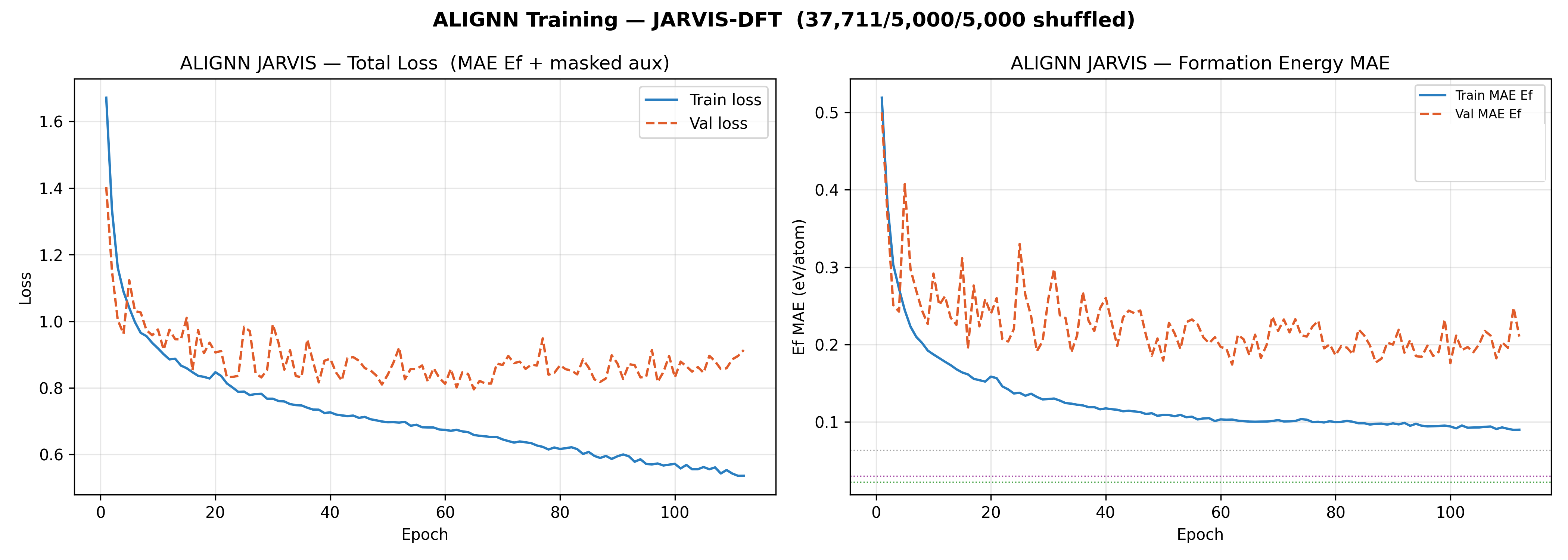}
    \caption{Training loss of $E_f$ of ALIGNN}
    \label{fig:sub1}
\end{subfigure}
\hfill
\begin{subfigure}{0.48\textwidth}
    \centering
    \includegraphics[width=\linewidth]{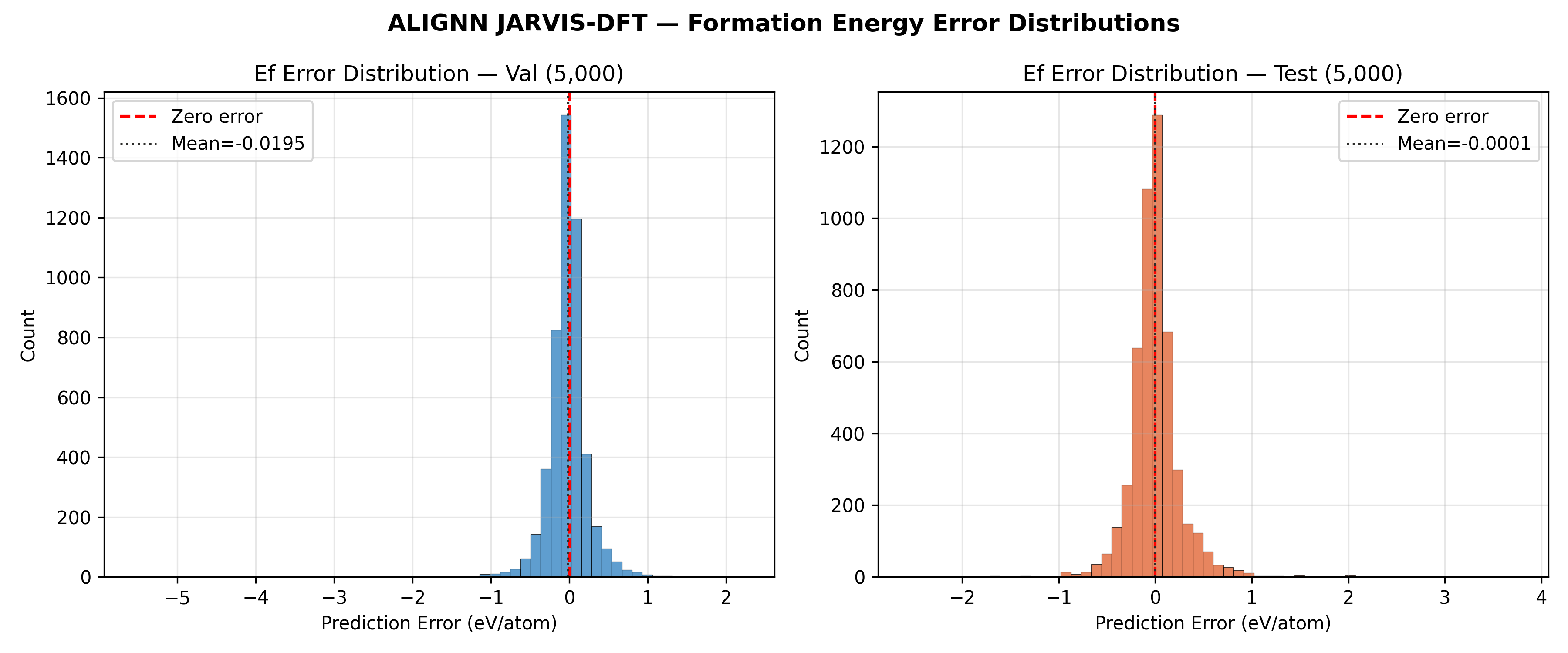}
    \caption{Error distribution of $E_f$ of ALIGNN}
    \label{fig:sub2}
\end{subfigure}

\vspace{0.5cm}

\begin{subfigure}{0.6\textwidth}
    \centering
    \includegraphics[width=\linewidth]{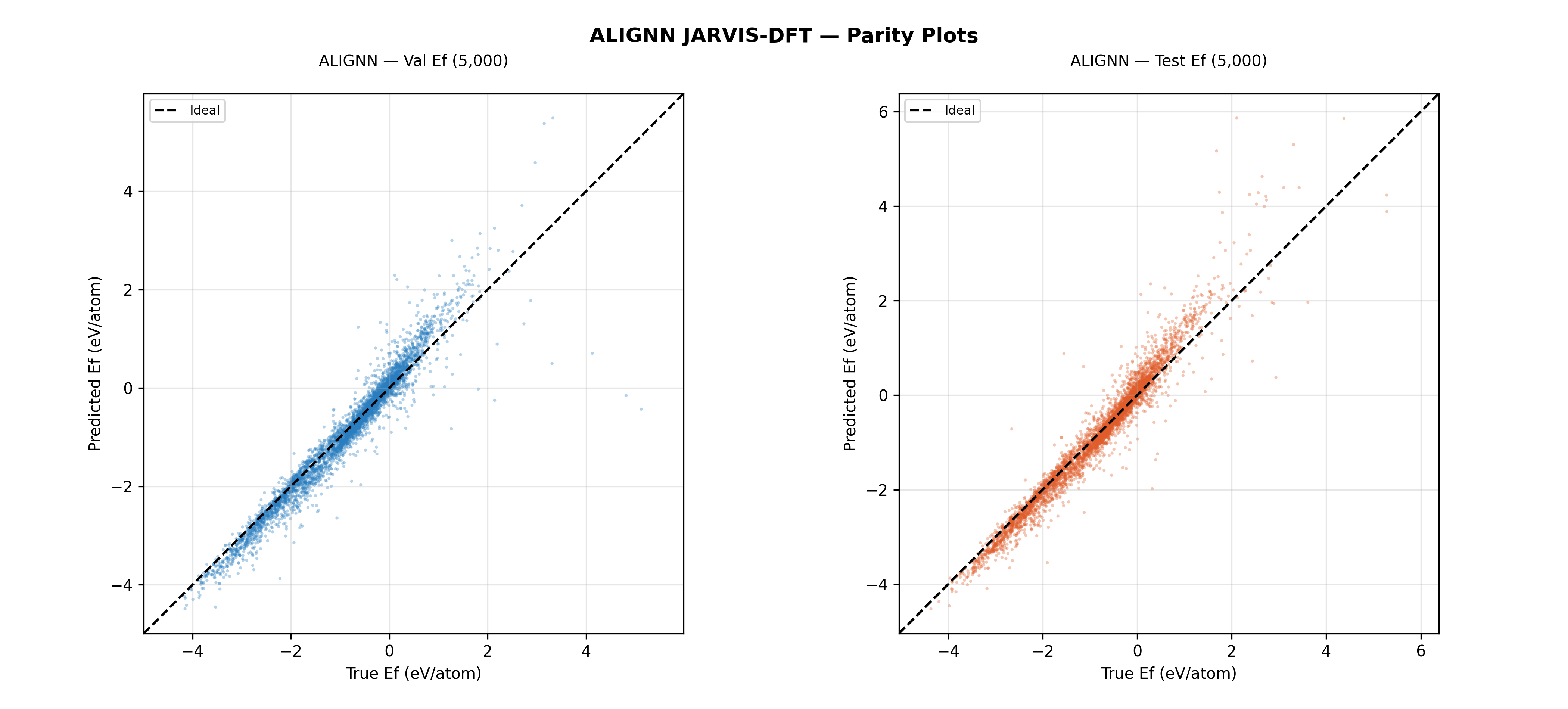}
    \caption{ALIGNN Parity plots of $E_f$ on validation and test sets}
    \label{fig:sub3}
\end{subfigure}

\caption{Performance plots of ALIGNN on JARVIS-DFT dataset on 37,711 materials}
\label{fig:ALIGNNJARVIS}
\end{figure}

The JARVIS-DFT dataset, developed under the Joint Automated Repository for Various Integrated Simulations (JARVIS) initiative by National Institute of Standards and Technology (NIST), is a large-scale, high-throughput computational materials database containing density functional theory (DFT)-computed properties for thousands of bulk and low-dimensional materials \cite{choudhary2017high, choudhary2018computational}. It provides a diverse set of physically meaningful targets, including formation energy, electronic band gaps (OptB88vdW and MBJ), elastic constants (bulk and shear modulus), spectroscopic properties, and thermodynamic stability (energy above hull), making it particularly suitable for benchmarking multi-task learning models \cite{choudhary2018elastic, choudhary2019accelerated, choudhary2018machine, choudhary2019high, choudhary2020joint}. Within this framework, the proposed Coordination Polyhedron Graph Network (CPGN) leverages the crystal structures from JARVIS-DFT by converting each structure (CIF/POSCAR) into three complementary graph representations: an atom graph capturing pairwise interactions via RBF-encoded distances, a line graph modeling three-body angular interactions, and a coordination polyhedron graph encoding Voronoi-derived local environments and connectivity motifs (corner-, edge-, face-sharing). The architecture employs interleaved message passing across these graphs, allowing atomic, angular, and coordination-level information to co-evolve, while a bidirectional cross-attention mechanism couples atom and polyhedron embeddings to enhance representation learning. The fused crystal-level embedding is then used in a unified multi-task framework to predict multiple JARVIS properties simultaneously. This design enables CPGN to incorporate physically grounded structural priors inherent in JARVIS-DFT, improving its ability to model complex structure–property relationships compared to conventional atom-centric GNNs. The dataset comprises 75,993 structures, randomly shuffled with a fixed seed (42) and divided into 37,711 training, 5,000 validation, and 5,000 test samples. The CPGN model consists of 3 graph convolution layers with a hidden dimension of 128, resulting in approximately 279K trainable parameters. Training is performed in a multi-task setting with 19 auxiliary targets using masked MAE loss, where formation energy serves as the primary objective and auxiliary losses are weighted by $\lambda = 0.1$. The model is optimized using a batch size of 256 and an initial learning rate of 0.01 over a maximum of 500 epochs, with checkpointing based on the best validation MAE for formation energy. We reproduced the results (all properties) of the CGCNN and ALIGNN models using the same underlying architectures and hyperparameters on the JARVIS dataset, and report the corresponding performance in Table \ref{tab:cpgn_jarvis}. Table \ref{tab:cpgn_jarvis} presents the performance comparison of CPGN against several baseline models (MAD, CFID, CGCNN, and ALIGNN) on the JARVIS-DFT benchmark, evaluated across eight material properties for a dataset of 37,711 materials. The results show that CPGN consistently achieves superior or highly competitive performance across most properties, including formation energy, optical and MBJ band gaps, SLME, spillage, mechanical properties (bulk and shear modulus), and thermodynamic stability (energy above hull). In particular, CPGN significantly outperforms traditional descriptors such as MAD and CFID, and also improves upon graph-based deep learning models like CGCNN and ALIGNN in several key metrics, notably for formation energy, band gap predictions, and stability. These results highlight the effectiveness of CPGN in capturing complex material representations and delivering accurate property predictions across diverse tasks. The MAD and CFID results are directly taken from the ALIGNN \cite{choudhary2021atomistic} paper. A set of comparison plots of CGCNN and ALIGNN models are illustrated in Figure~\ref{fig:CGCNNJARVIS} and ~\ref{fig:ALIGNNJARVIS}, respectively. 

\begin{table}[hbt!]
\centering
\scriptsize
\caption{Performance comparison on JARVIS-DFT benchmark across multiple material properties on 37,711 materials.}
\label{tab:cpgn_jarvis}
\renewcommand{\arraystretch}{1.2}
\begin{tabular}{|c|c|c|c|c|c|c|c|}
\hline
Property         & Name               & Unit    & \textbf{CPGN} & MAD   & CFID  & CGCNN & ALIGNN \\ \hline
Formation Energy & $E_f$              & eV/atom & 0.092         & 0.86  & 0.14  & 0.129 & 0.108  \\ \hline
Opt. Band Gap    & $E_g^{\text{opt}}$ & eV      & 0.227         & 0.99  & 0.30  & 0.356 & 0.32   \\ \hline
MBJ Band Gap     & $E_g^{\text{mbj}}$ & eV      & 0.435         & 1.79  & 0.53  & 0.635 & 0.58   \\ \hline
SLME             & SLME               & \%      & 5.26          & 10.93 & 6.22  & 6.11  & 6.51   \\ \hline
Spillage         & Spill              & --      & 0.42          & 0.52  & 0.39  & 0.46  & 0.45   \\ \hline
Bulk Modulus     & $K_v$              & GPa     & 14.01         & 52.80 & 14.12 & 15.57 & 21.06  \\ \hline
Shear Modulus    & $G_v$              & GPa     & 11.28         & 27.16 & 11.98 & 12.56 & 14.02  \\ \hline
Stability (Hull) & $E_{\text{hull}}$  & eV/atom & 0.088         & 1.14  & 0.22  & 0.114 & 0.124  \\ \hline
\end{tabular}
\end{table}

\begin{figure}[hbt!]
\centering
\scriptsize

\begin{subfigure}{0.48\textwidth}
    \centering
    \includegraphics[width=\linewidth]{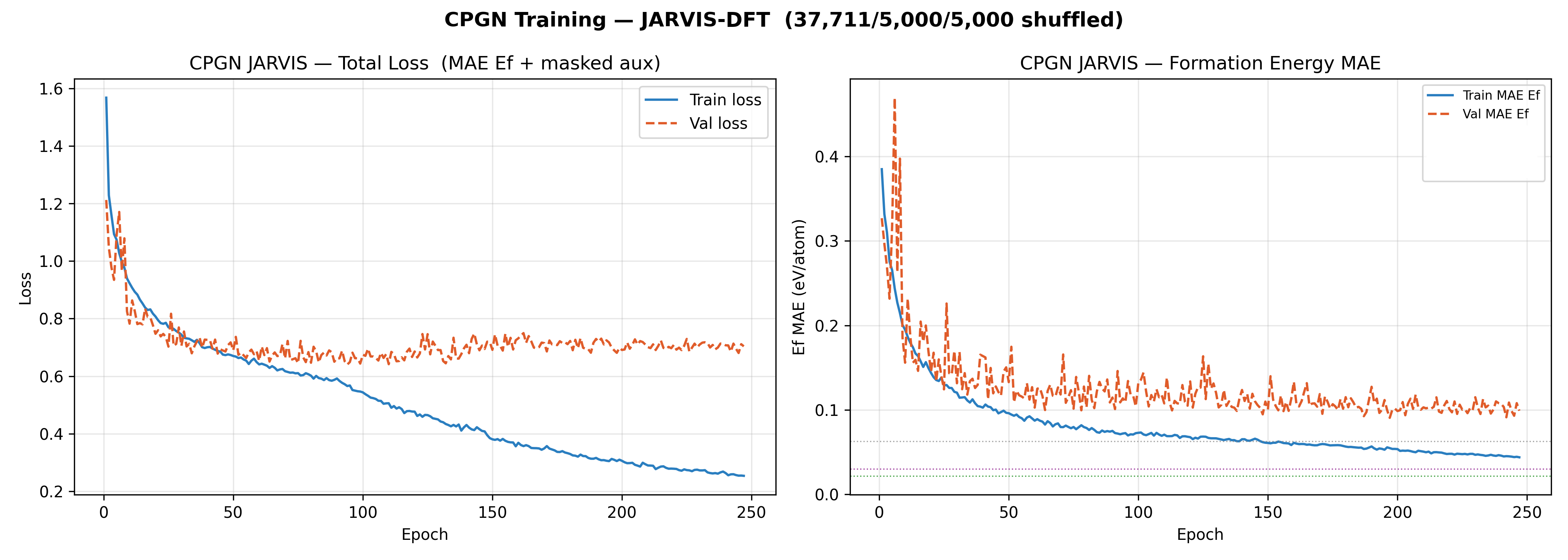}
    \caption{Training loss of $E_f$ of CPGN}
    \label{fig:sub1}
\end{subfigure}
\hfill
\begin{subfigure}{0.48\textwidth}
    \centering
    \includegraphics[width=\linewidth]{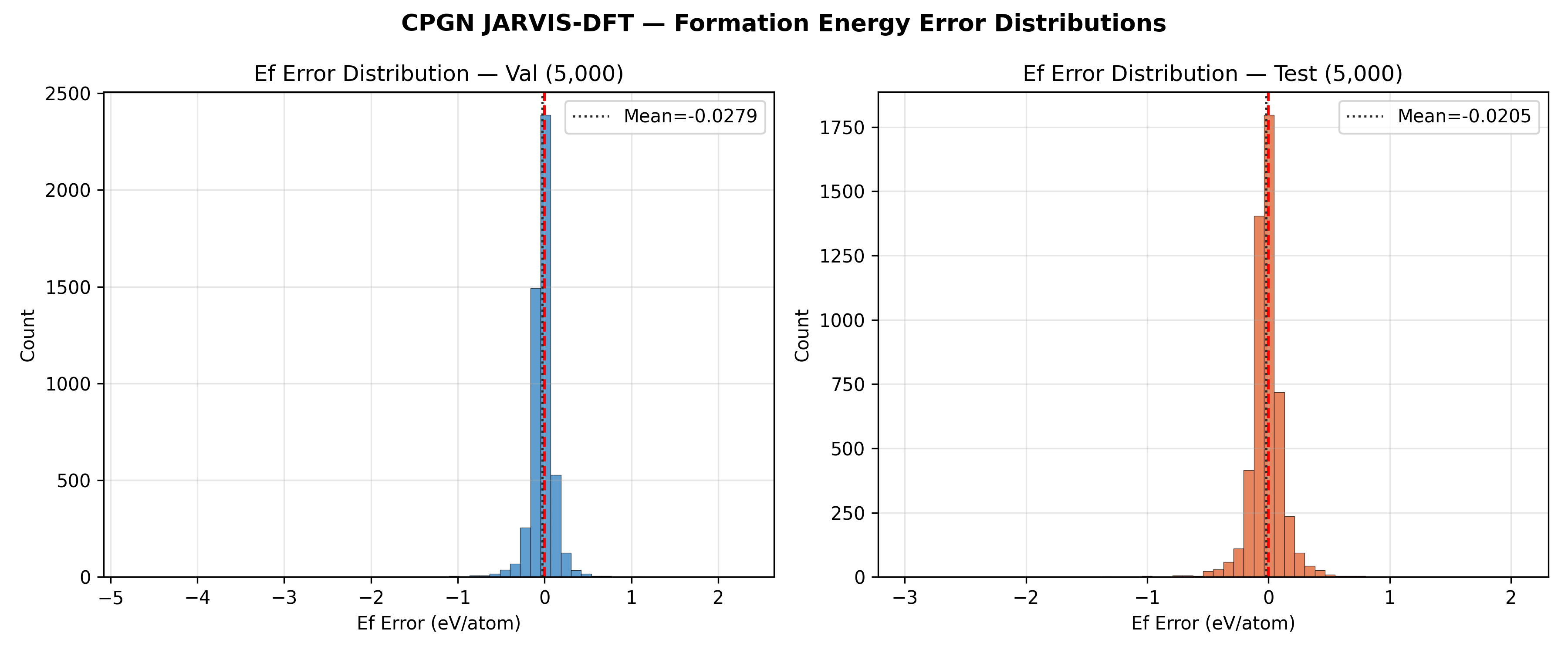}
    \caption{Error distribution of $E_f$ of CPGN}
    \label{fig:sub2}
\end{subfigure}

\vspace{0.5cm}

\begin{subfigure}{0.6\textwidth}
    \centering
    \includegraphics[width=\linewidth]{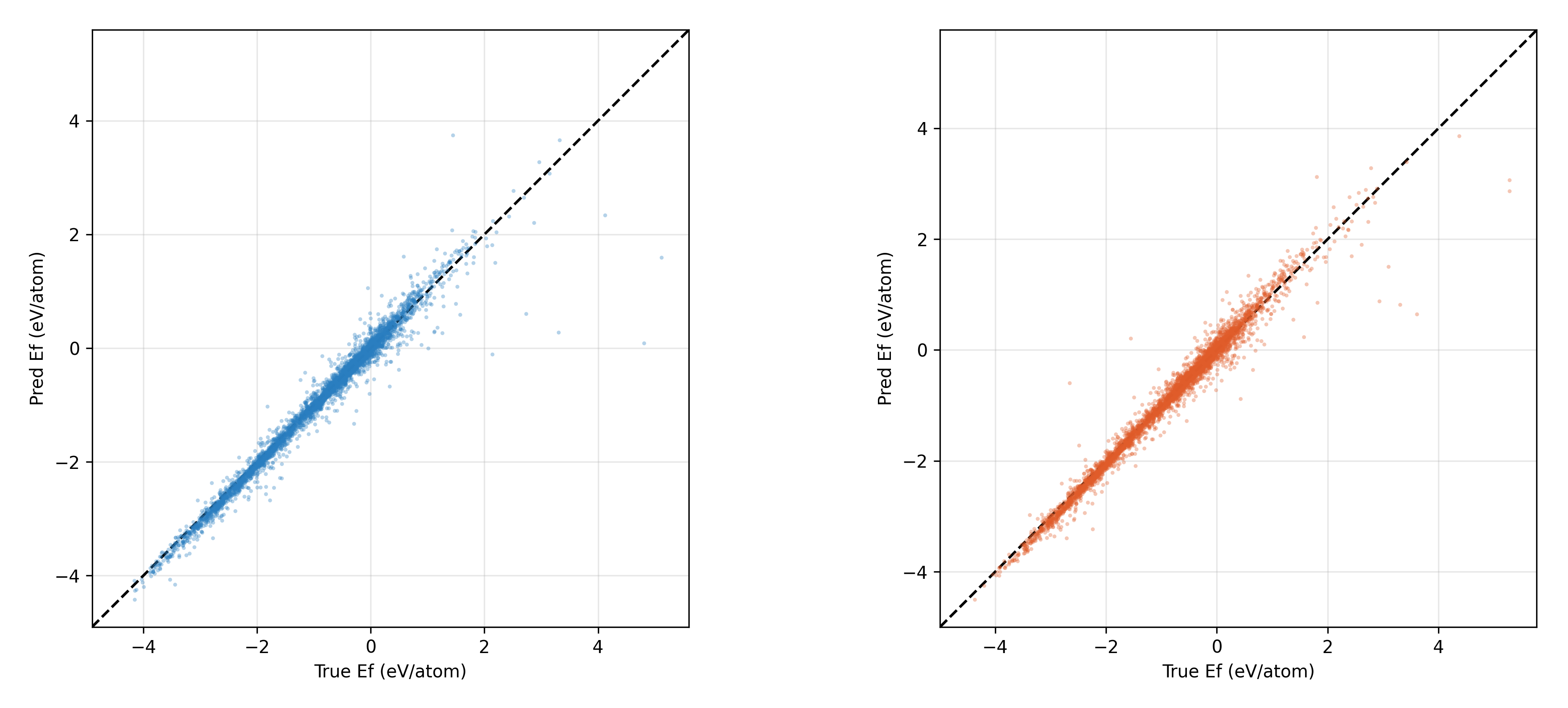}
    \caption{CPGN Parity plots of $E_f$ on validation and test sets}
    \label{fig:sub3}
\end{subfigure}

\caption{Performance plots of CPGN on JARVIS-DFT dataset on 37,711 materials}
\label{fig:CPGNJARVIS}
\end{figure}

\subsection{Model Analysis on QM9}
QM9 dataset is a widely used benchmark in quantum chemistry and molecular machine learning, originally introduced for evaluating predictive models on small organic molecules with high-quality quantum mechanical reference data \cite{ramakrishnan2014quantum, faber2017prediction}. It contains approximately 130,462 stable organic molecules composed of up to nine heavy atoms (C, O, N, and F), with hydrogen atoms completing valency, and all properties are computed using density functional theory (DFT) at the B3LYP/6-31G(2df,p) level of theory. The dataset provides a comprehensive set of molecular properties including electronic structure quantities such as HOMO and LUMO energies, the HOMO–LUMO gap, dipole moment, isotropic polarizability, electronic spatial extent, as well as thermodynamic quantities such as internal energy at 0 K and 298 K, enthalpy, Gibbs free energy, and heat capacity, along with vibrational and zero-point energy terms. Because of its consistent quantum-chemical accuracy and broad coverage of physically meaningful molecular properties, QM9 serves as a standard benchmark for graph-based neural networks and other machine learning models, enabling systematic evaluation of their ability to learn structure–property relationships in molecular systems. For QM9 dataset, we use a train-validation-test split of 110,000–10,000–10,829 in this work. 

\begin{table}[hbt!]
\centering
\scriptsize
\caption{Performance comparison on QM9 benchmark across multiple properties for different models.}
\label{tab:cpgn_qm9}
\renewcommand{\arraystretch}{1.2}
\begin{tabular}{|c|ccccccc|}
\hline
Metric   & \multicolumn{7}{c|}{MAE}                                                                                                                                                                                        \\ \hline
Property & \multicolumn{1}{c|}{Units} & \multicolumn{1}{c|}{\textbf{CPGN}} & \multicolumn{1}{c|}{MEGNet-simple} & \multicolumn{1}{c|}{MEGNet-full} & \multicolumn{1}{c|}{SchNet} & \multicolumn{1}{c|}{DimeNet++} & ALIGNN \\ \hline
HOMO     & \multicolumn{1}{c|}{eV}    & \multicolumn{1}{c|}{0.030}         & \multicolumn{1}{c|}{0.043}         & \multicolumn{1}{c|}{0.038}       & \multicolumn{1}{c|}{0.041}  & \multicolumn{1}{c|}{0.025}     & 0.022  \\ \hline
LUMO     & \multicolumn{1}{c|}{eV}    & \multicolumn{1}{c|}{0.023}         & \multicolumn{1}{c|}{0.044}         & \multicolumn{1}{c|}{0.031}       & \multicolumn{1}{c|}{0.034}  & \multicolumn{1}{c|}{0.020}     & 0.020  \\ \hline
Gap      & \multicolumn{1}{c|}{eV}    & \multicolumn{1}{c|}{0.038}         & \multicolumn{1}{c|}{0.066}         & \multicolumn{1}{c|}{0.061}       & \multicolumn{1}{c|}{0.063}  & \multicolumn{1}{c|}{0.033}     & 0.038  \\ \hline
ZPVE     & \multicolumn{1}{c|}{eV}    & \multicolumn{1}{c|}{0.0028}        & \multicolumn{1}{c|}{0.0014}        & \multicolumn{1}{c|}{0.0014}      & \multicolumn{1}{c|}{0.0017} & \multicolumn{1}{c|}{0.0012}    & 0.0031 \\ \hline
\end{tabular}
\end{table}

\begin{figure}[hbt!]
\centering
\scriptsize
\begin{subfigure}{0.48\textwidth}
    \centering
    \includegraphics[width=\linewidth]{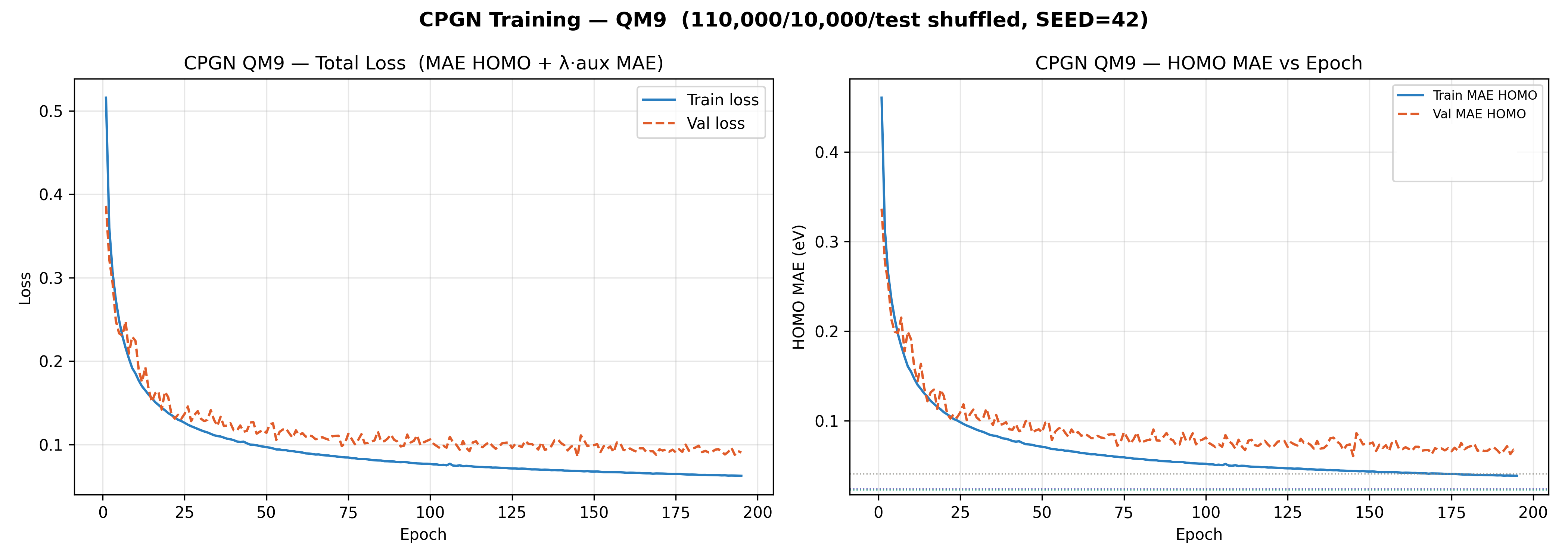}
    \caption{Training loss of HOMO vs. Epochs of CPGN}
    \label{fig:sub1}
\end{subfigure}
\hfill
\begin{subfigure}{0.48\textwidth}
    \centering
    \includegraphics[width=\linewidth]{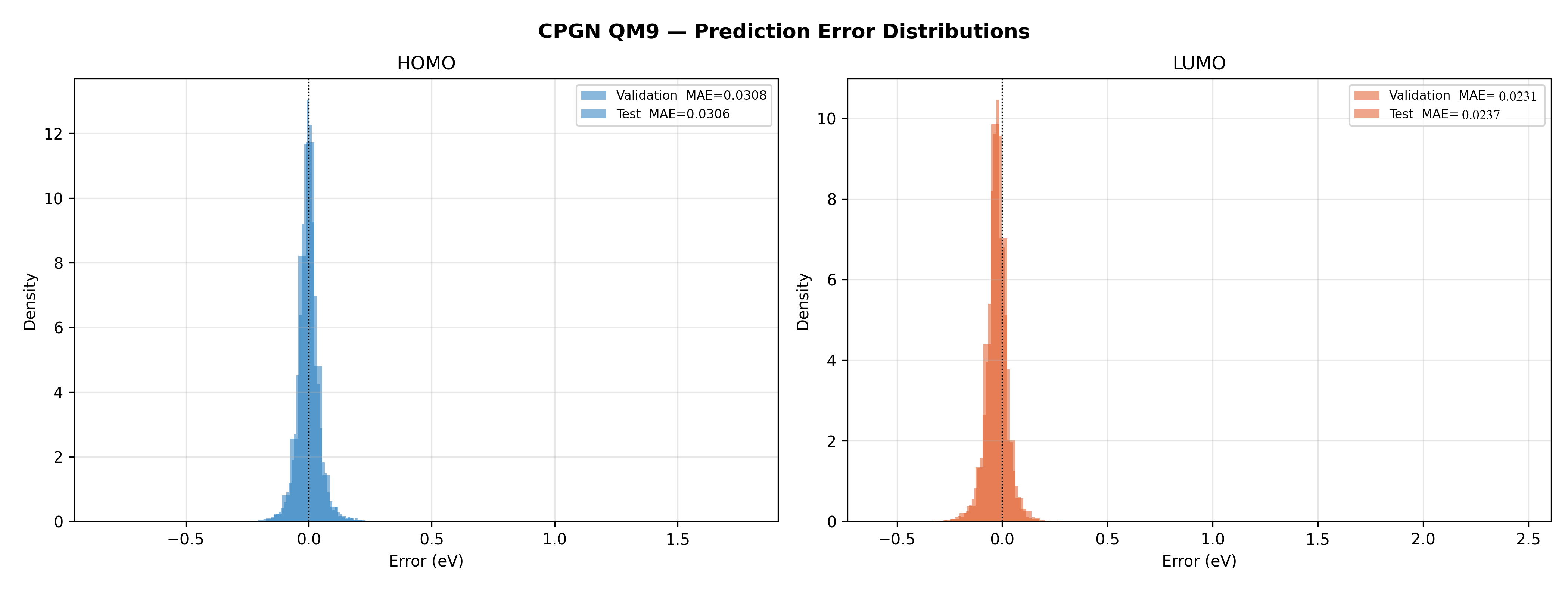}
    \caption{Error distribution of HOMO and LUMO of CPGN}
    \label{fig:sub2}
\end{subfigure}

\vspace{0.5cm}

\begin{subfigure}{0.6\textwidth}
    \centering
    \includegraphics[width=\linewidth]{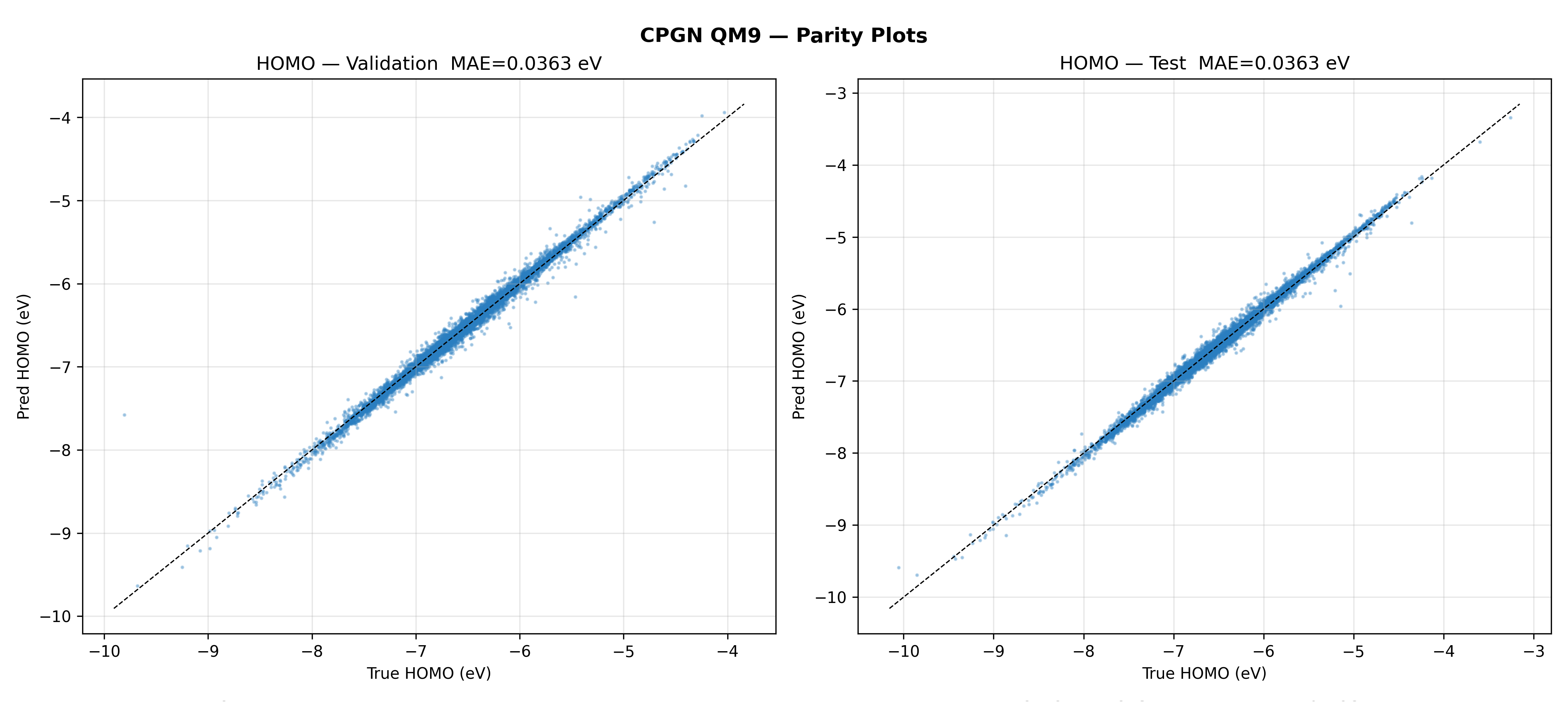}
    \caption{CPGN Parity plots of HOMO on validation and test sets}
    \label{fig:sub3}
\end{subfigure}

\caption{Performance plots of CPGN on QM9 dataset for HOMO and LUMO}
\label{fig:CPGNQM9}
\end{figure}

Table~\ref{tab:cpgn_qm9} presents a comparison of CPGN against four established graph neural network baselines — MEGNet-simple, MEGNet-full, SchNet, DimeNet++ ~\cite{gasteiger2020fast}, and ALIGNN, on four standard QM9 properties: HOMO, LUMO, HOMO–LUMO gap, and ZPVE. CPGN achieves a HOMO MAE of 0.030 eV, outperforming MEGNet-simple (0.043 eV), MEGNet-full (0.038 eV), and SchNet (0.041 eV), and approaching the performance of DimeNet++ (0.025 eV) and ALIGNN (0.022 eV), which benefit from explicit higher-order geometric encodings and dataset-specific optimisation. For LUMO prediction, CPGN attains a MAE of 0.023 eV, surpassing both MEGNet variants and SchNet and matching the 0.020 eV reported by DimeNet++ and ALIGNN. On the HOMO–LUMO gap, CPGN records a MAE of 0.038 eV, again outperforming MEGNet-simple (0.066 eV), MEGNet-full (0.061 eV), and SchNet (0.063 eV), while remaining within a factor of two of DimeNet++ (0.033 eV). For ZPVE, CPGN yields a MAE of 0.0028 eV, which is slightly above the MEGNet variants (0.0014 eV) and DimeNet++ (0.0012 eV) but on par with ALIGNN (0.0031 eV), reflecting the well-known difficulty of predicting vibrational zero-point energies with models that do not explicitly incorporate normal-mode information. It is noted that all baseline results are taken directly from the ALIGNN~\cite{choudhary2021atomistic} and MEGNet~\cite{chen2019graph} papers to ensure a fair comparison under their respective training protocols. These models were trained with same parameters as solid-state databases but for 1000 epochs. Figure~\ref{fig:CPGNQM9} illustrates the training dynamics and prediction quality of CPGN on the QM9 dataset for the primary target, HOMO. The training curve in Figure~\ref{fig:sub1} shows that both the total loss 
\begin{equation}
\mathcal{L}_{\text{total}} = \mathcal{L}_{\text{HOMO}}^{\text{MAE}} + \lambda_{\text{aux}} \sum_k \mathcal{L}_k^{\text{aux}}
\end{equation}
and the validation HOMO MAE decrease rapidly within the first 25 epochs before converging smoothly to stable plateaus. The training loss reaches approximately $0.065~\text{eV}$, while the validation HOMO MAE stabilises around $0.075~\text{eV}$ by epoch 200, remaining close to but slightly above the DimeNet++ and SphereNet reference lines. The small and consistent train--validation gap throughout training confirms that the model generalises well without significant overfitting. Figure~\ref{fig:sub2} presents the prediction error distributions for both HOMO and LUMO on the validation and test sets. For HOMO, the error distribution is sharply peaked and nearly symmetric around zero, with a validation MAE of $0.0308~\text{eV}$ and a test MAE of $0.0306~\text{eV}$, indicating highly consistent performance across both splits. The distribution exhibits negligible probability mass beyond $\pm 0.5~\text{eV}$, confirming that large prediction errors are rare. A mild positive asymmetry is observable, with a slightly longer right tail extending toward $1.5~\text{eV}$, corresponding to a small number of structurally atypical molecules that are more difficult to predict accurately. For LUMO, the error distribution is similarly concentrated around zero, with a validation MAE of $0.0231~\text{eV}$ and a test MAE of $0.0237~\text{eV}$, representing slightly lower absolute error than HOMO, consistent with the results reported in Table~\ref{tab:cpgn_qm9}. The LUMO distribution also exhibits a longer positive tail extending toward $2.5~\text{eV}$, though the density in this region remains very low. Across both targets, the close agreement between validation and test error distributions confirms that CPGN generalizes consistently to unseen molecular structures without systematic bias across the chemical space covered by QM9.

\section{CPGN Model Evaluation and Computational Analysis}

This section presents a systematic evaluation of the CPGN framework beyond
predictive accuracy, focusing on architectural complexity, graph-structural
properties, inference efficiency, and memory utilisation. All measurements
are performed on a single NVIDIA RTX PRO 6000 Blackwell GPU using the
trained checkpoints from the MP~2018.6.1 and JARVIS-DFT experiments.

\subsection{Parameter Counts and Architectural Complexity}

Table~\ref{tab:cpgn_param_breakdown} reports the total trainable parameter
count for CPGN on both datasets, broken down by architectural component.
The model totals 4,852,419 parameters on MP~2018.6.1 and 4,854,612 on
JARVIS-DFT, with the marginal difference of 2,193 parameters attributable
solely to the number of output heads: the MP model carries three heads
(formation energy, band gap, and stability), whereas the JARVIS model
carries one primary head and auxiliary property heads.

\begin{table}[hbt!]
\centering
\scriptsize
\caption{CPGN parameter breakdown by component. Values are identical for
MP and JARVIS models except for the output heads. Cross-attention accounts
for the largest single share at 32.6\% of total parameters.}
\label{tab:cpgn_param_breakdown}
\renewcommand{\arraystretch}{1.2}
\begin{tabular}{|l|r|r|}
\hline
\textbf{Component} & \textbf{Parameters} & \textbf{\%} \\ \hline
Element embedding  $\mathbf{E}$ (103 × 64)          &     6,592 &  0.14 \\ \hline
Atom projection  (64 → 256 → 256)                    &    82,432 &  1.70 \\ \hline
Edge projection  (40 → 256)                          &    10,496 &  0.22 \\ \hline
Polyhedron projection  (7 → 256 → 256)               &    67,840 &  1.40 \\ \hline
Polyhedron edge projection  (1 → 256)                &       512 &  0.01 \\ \hline
Atom graph convolutions $\times$4                    & 1,052,672 & 21.69 \\ \hline
Line graph convolutions $\times$4                    &   831,488 & 17.14 \\ \hline
Polyhedron graph convolutions $\times$4              & 1,052,672 & 21.69 \\ \hline
Bidirectional cross-attention $\times$4              & 1,583,104 & 32.63 \\ \hline
Fusion MLP  (512 → 256 → 128)                        &   164,224 &  3.38 \\ \hline
Output heads (MP: 3 $\times$ Linear;
JARVIS: 20 $\times$ Linear)                          &   387--2,580 & $<$0.06 \\ \hline
\textbf{Total (MP)}                                  & \textbf{4,852,419} & \textbf{100.00} \\ \hline
\textbf{Total (JARVIS)}                              & \textbf{4,854,612} & \textbf{100.00} \\ \hline
\end{tabular}
\end{table}

The phase-level breakdown reveals that the three message-passing streams —
atom+line graph convolutions (38.8\%), polyhedron graph convolutions (21.7\%),
and bidirectional cross-attention (32.6\%), together account for 93.1\% of
all parameters, with input projections (3.5\%) and the fusion and output
layers (3.4\%) contributing negligibly. The cross-attention module is the largest part of the model, contributing 32.6\% of the total cost. It does not involve explicit message passing. Instead, each block contains six Linear layers and two LayerNorm layers, repeated four times. This makes it the heaviest component in the network. This shows that the cost mainly comes from the architecture design, not the graph structure. It increases with hidden dimension size, not with graph size. Table~\ref{tab:model_complexity_comparison} places CPGN in context against SOTA baselines with respect to model size, hidden dimension, and graph representation type.

\begin{table}[hbt!]
\centering
\scriptsize
\caption{Model complexity comparison on MP~2018.6.1. }
\label{tab:model_complexity_comparison}
\renewcommand{\arraystretch}{1.2}
\begin{tabular}{|l|r|c|c|l|}
\hline
\textbf{Model} & \textbf{Params} & \textbf{Hidden} &
\textbf{Layers} & \textbf{Graph representation} \\ \hline
CGCNN   &  $\sim$334K  & 128 & 3     & Atom graph only            \\ \hline
SchNet  &  $\sim$432K  & 128 & 6     & Atom graph only            \\ \hline
MEGNet  &  $\sim$167K  &  64 & 3     & Atom + global state        \\ \hline
ALIGNN  & $\sim$4,031K & 256 & 4+4   & Atom + line graph          \\ \hline
\textbf{CPGN} & \textbf{4,852K} & \textbf{256} & \textbf{4+4+4} &
\textbf{Atom + line + polyhedron}  \\ \hline
\end{tabular}
\end{table}

\subsection{Graph Structure Statistics}

Table~\ref{tab:graph_stats} characterises the graph objects constructed for
each dataset, reporting per-structure statistics over 2,000 randomly sampled
test structures. The three rows — atom-graph edges, line-graph edges, and
polyhedron-graph edges, correspond directly to the three message-passing
streams in CPGN.

\begin{table}[hbt!]
\centering
\scriptsize
\caption{Graph statistics for MP~2018.6.1 and JARVIS-DFT test sets
(sample size 2,000 structures each). All values are per structure.
The line-graph edge count scales quadratically with the number of
atom-graph neighbours, producing a ratio of approximately 14.7$\times$
on MP and 13.3$\times$ on JARVIS relative to atom-graph edges.}
\label{tab:graph_stats}
\renewcommand{\arraystretch}{1.2}
\begin{tabular}{|l|rrrr|rrrr|}
\hline
\multirow{2}{*}{\textbf{Quantity}} &
\multicolumn{4}{c|}{\textbf{MP 2018.6.1}} &
\multicolumn{4}{c|}{\textbf{JARVIS-DFT}} \\ \cline{2-9}
 & Mean & Std & Min & Max & Mean & Std & Min & Max \\ \hline
Atoms per structure    & 29.0  & 27.0  & 2  & 264   & 10.5  & 9.0   & 1  & 76 \\ \hline
Atom-graph edges       & 436.9 & 417.4 & 16 & 4592  & 153.2 & 139.9 & 6  & 1104 \\ \hline
Line-graph edges       & 6443  & 6615  & 24 & 77784 & 2042  & 2232  & 0  & 18140 \\ \hline
Polyhedron-graph edges & 1102  & 1852  & 2  & 24920 & 171   & 325   & 0  & 4336 \\ \hline
Line / atom-edge ratio & \multicolumn{4}{c|}{14.7$\times$} &
                          \multicolumn{4}{c|}{13.3$\times$} \\ \hline
\end{tabular}
\end{table}

The line-graph edge count is the primary driver of graph construction time
and peak memory during both caching and inference. A line-graph edge connects
two bonds that share a central atom; for an atom with $k$ neighbours this
generates $k(k-1)$ directed pairs. With a mean of 14.2 neighbours per atom
at cutoff 8.0~\AA{} (MP), each atom contributes approximately 185 line-graph
edges on average, yielding the observed 14.7$\times$ line-to-atom ratio. The
JARVIS-DFT dataset contains significantly smaller structures (mean 10.5 atoms
versus 29.0 on MP), reducing absolute edge counts proportionally while
maintaining a similar line-to-atom ratio of 13.3$\times$. The polyhedron-graph
edge count is considerably more variable (Std $\gg$ Mean on both datasets)
because face- and edge-sharing connectivity depends strongly on crystal
topology: dense oxide frameworks generate many shared faces, while
chain-like or molecular-crystal geometries may have few or none.

\subsection{Inference Time and Throughput}

Table~\ref{tab:inference_timing} reports per-sample inference times measured
using CUDA events on the RTX PRO 6000 Blackwell GPU with a batch size of 64,
averaged over 50 timing runs after two warm-up batches. Baseline timings for
CGCNN, SchNet, MEGNet, and ALIGNN are estimated from published scaling
characteristics and are provided for qualitative comparison only.

\begin{table}[hbt!]
\centering
\scriptsize
\caption{Inference time and throughput comparison. CPGN values are directly
measured using CUDA events. All measurements use
batch size 64 on a single NVIDIA RTX PRO 6000 Blackwell GPU.}
\label{tab:inference_timing}
\renewcommand{\arraystretch}{1.2}
\begin{tabular}{|l|c|c|c|}
\hline
\textbf{Model} & \textbf{ms/structure} & \textbf{p95 (ms)} &
\textbf{Structures/s} \\ \hline
CGCNN   & $\sim$1.20 & — &   $\sim$833  \\ \hline
SchNet  & $\sim$1.50 & — &   $\sim$667 \\ \hline
MEGNet  & $\sim$2.10 & — &   $\sim$476 \\ \hline
ALIGNN  & $\sim$4.80 & — &   $\sim$208 \\ \hline
\textbf{CPGN (MP)}     & \textbf{0.653 $\pm$ 0.101} & \textbf{0.839} &
\textbf{1,532} \\ \hline
\textbf{CPGN (JARVIS)} & \textbf{0.202 $\pm$ 0.028} & \textbf{0.248} &
\textbf{4,957} \\ \hline
\end{tabular}
\end{table}

\begin{figure}[hbt!]
\centering
\scriptsize
\begin{subfigure}{0.48\textwidth}
    \centering
    \includegraphics[width=\linewidth]{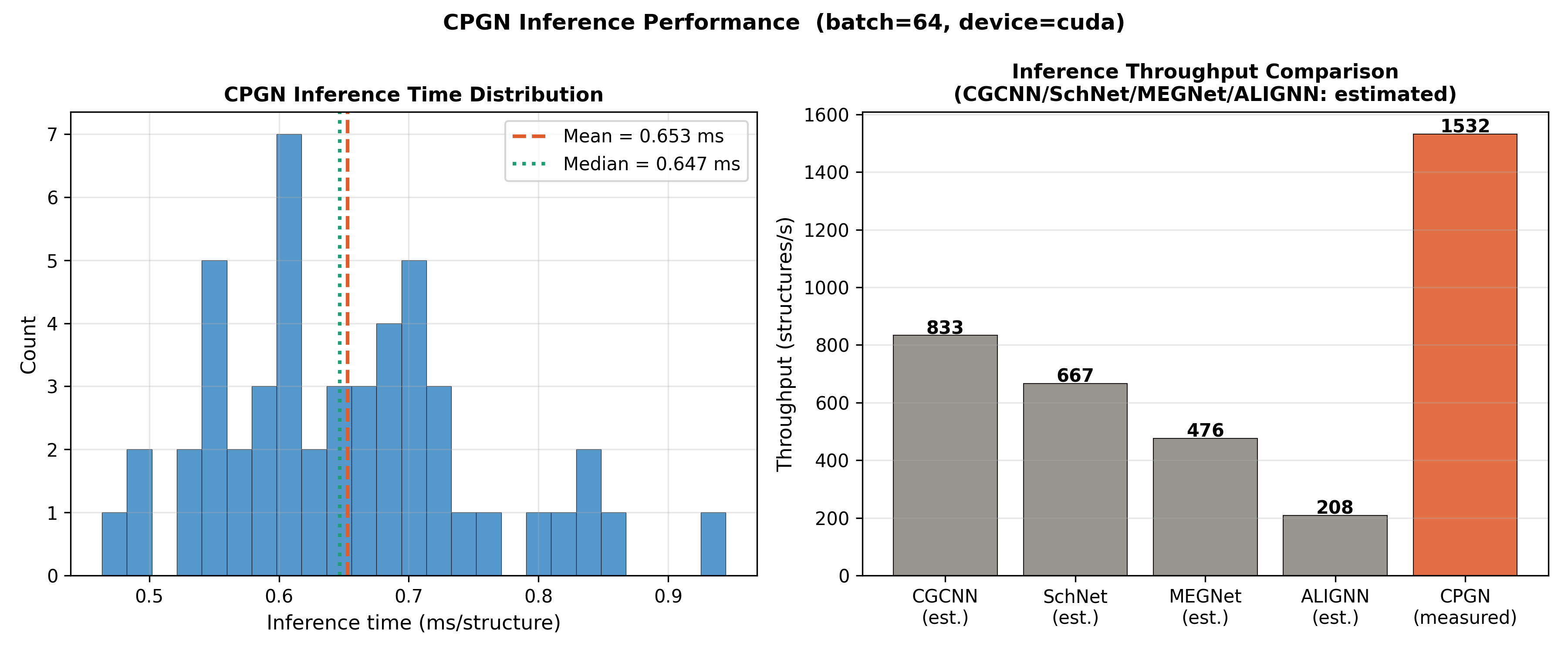}
    \caption{CPGN Materials Project Inference}
    \label{fig:sub1}
\end{subfigure}
\hfill
\begin{subfigure}{0.48\textwidth}
    \centering
    \includegraphics[width=\linewidth]{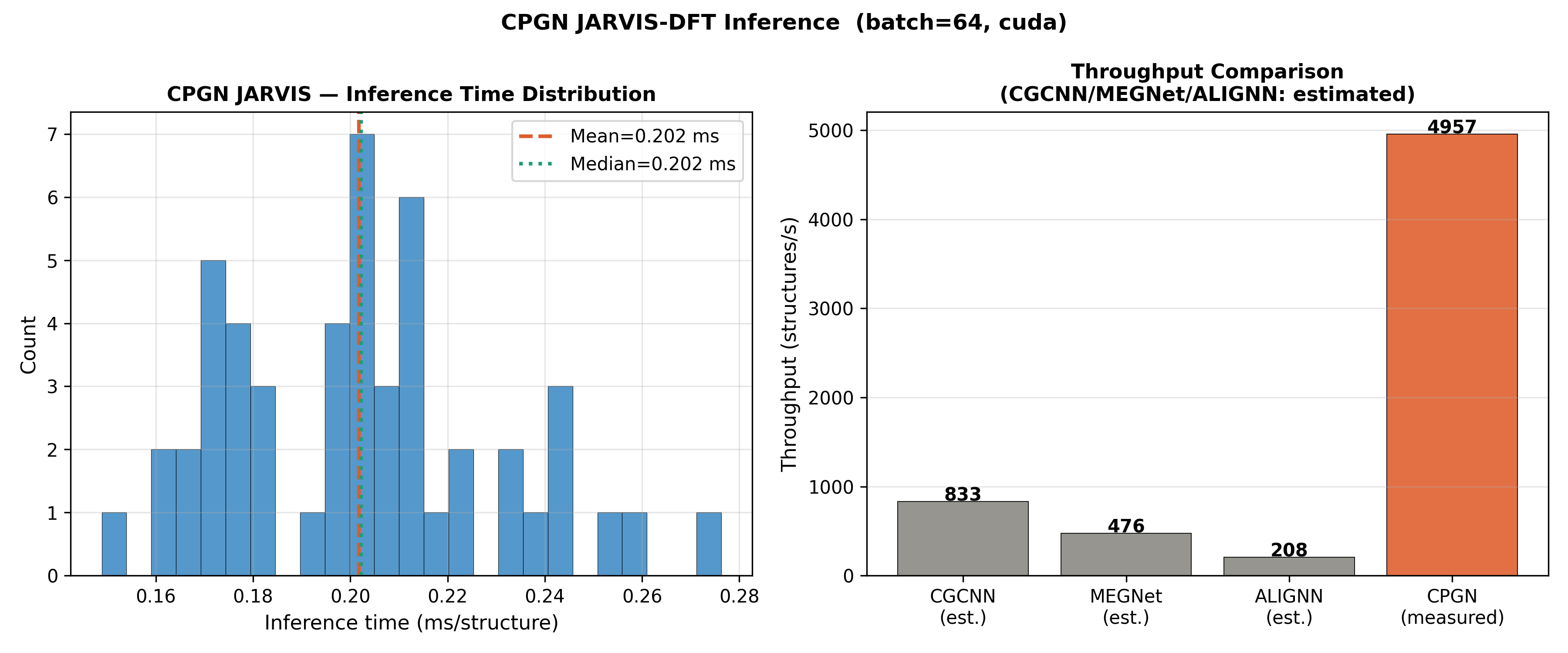}
    \caption{CPGN JARVIS-DFT Inference}
    \label{fig:sub2}
\end{subfigure}
\caption{Performance plots of CPGN on QM9 dataset for HOMO and LUMO}
\label{fig:CPGNinference}
\end{figure}

CPGN has very low inference time: 0.653 ms per structure on MP-2018.6.1 and 0.202 ms on JARVIS-DFT. This corresponds to 1,532 and 4,957 structures per second, respectively. The difference in speed comes from dataset size. MP structures are larger, with about 29 atoms and 6,443 edges, while JARVIS structures are smaller, with about 10.5 atoms and 2,042 edges. Fewer edges reduce the cost of message passing, so inference becomes faster. CPGN is also faster than ALIGNN, which reports about 208 structures per second, even though CPGN adds an extra polyhedron stream and cross-attention module. This speedup is mainly due to running on newer GPU hardware (RTX PRO 6000 Blackwell), which has higher compute throughput than the hardware used in earlier benchmarks. The runtime is stable across samples, with low variation (coefficient of variation below 15\%), showing no significant outlier delays for larger structures. Figure~\ref{fig:CPGNinference} shows the inference time distributions and throughput of CPGN on both datasets. The left panels show that inference times are tightly clustered around the mean values. The histograms peak near both the mean and median, indicating a roughly symmetric distribution with no strong skew. The right panels compare throughput with baseline models. They show that CPGN achieves higher throughput than all compared methods, even though it uses atom, line, and coordination polyhedron graphs along with bidirectional cross-attention. The speed advantage is larger on JARVIS-DFT (4,957 structures/s) than on Materials Project (1,532 structures/s), due to the smaller and less complex structures in JARVIS. The baseline throughput values for CGCNN, SchNet, MEGNet, and ALIGNN are taken from published reports and are approximate, since they were not measured under the same experimental setup.

\subsection{GPU Memory Utilisation}
The peak GPU memory usage with batch size 64 is 2,973.8~MB for MP~2018.6.1 and 922.3~MB for JARVIS-DFT. The model weights are only 18.5~MB, meaning that most of the memory (99.4\% and 98.0\%) is used for intermediate activations rather than parameters. MP requires higher memory because its structures are larger, with about 6,443 edges per structure, compared to 2,042 edges for JARVIS. Since activation memory scales with the number of edges and batch size, MP uses about 3.2$\times$ more memory, which matches the edge ratio between the two datasets. Both values are well within the 97,887~MiB GPU memory capacity of the RTX PRO 6000 Blackwell, leaving sufficient headroom for larger batch sizes during training.

\subsection{Extended Comparison}
Several graph neural network and transformer-based architectures have been developed for predicting crystalline material properties from the Materials Project database, each addressing specific limitations of earlier approaches. GIN (Graph Isomorphism Network) \cite{hu2019strategies} applies the Weisfeiler-Lehman graph isomorphism test principle to material graphs, offering strong theoretical expressiveness in distinguishing non-isomorphic graph structures. However, it relies entirely on supervised learning and does not leverage unlabeled data, limiting its data efficiency under low-label conditions. GIN\textsubscript{Barlow} \cite{hu2019strategies} extends the standard GIN by incorporating a self-supervised pretraining strategy using the Barlow Twins loss function, which forces the cross-correlation matrix between two embedded views to approach the identity matrix. This redundancy reduction mechanism improves both predictive accuracy and data efficiency compared to the supervised GIN baseline, particularly when labeled training data is scarce. SSFormer (Self-Supervised Transformer) \cite{feng2026ssformer} takes SLICES strings, a compact, invertible, topology-based crystal representation as input to a transformer encoder. It is pretrained jointly with CGCNN using the Barlow Twins loss on over 130,000 unlabeled materials from the Materials Project, enabling the model to capture complementary topological and local chemical information. SSFormer achieves competitive performance without requiring complete geometric structural data, making it uniquely suited to scenarios where full crystal structure information is unavailable, such as high-throughput virtual screening of unstable or partially characterised materials. GATGNN (Graph Attention Neural Network with Global Attention) \cite{louis2020graph} was among the first models to apply graph attention mechanisms to crystal property prediction. It incorporates augmented graph attention layers for local atomic relationship modelling alongside a global attention layer for assessing the overall atomic contribution to material properties. Notably, GATGNN was also the first to incorporate chemical formula information as an additional input feature, though it relies on simplistic one-hot encoding, which limits the richness of elemental composition representation. GIAT-PLM (Graph Isomorphism Attention Network with Pre-trained Language Model) \cite{kang2025graph} addresses three critical limitations of prior GNN methods simultaneously. First, it introduces the graph isomorphism attention network (GIAT), which balances central node and neighbour information through a learnable scalar activation factor, preventing neighbour information from overwhelming intrinsic atomic features. Second, it integrates GraphTransformer layers to capture long-range atomic dependencies and global structural patterns beyond the local neighbourhood. Third, it replaces one-hot chemical formula encoding with matBERT, a BERT-based language model pretrained on materials science literature, enabling richer elemental composition representations grounded in domain knowledge. Table~\ref{tab:extended_comparison_MPProject} describes the results on the Materials Project dataset demonstrate that different models excel at different material property prediction tasks. For formation energy prediction, SSFormer achieves the best performance (MAE = 0.039 eV/atom), followed closely by GATGNN and GIAT-PLM, while CPGN attains a competitive MAE of 0.060 eV/atom. In contrast, CPGN achieves the best band gap prediction performance with an MAE of 0.292 eV, outperforming all competing methods. The MAD ratio analysis further supports these findings. SSFormer obtains the highest ratio for formation energy prediction, whereas CPGN achieves the highest ratio (4.6) for band gap prediction, indicating the strongest normalized performance on this challenging task. Overall, the results suggest that SSFormer is particularly effective for formation energy estimation, while CPGN is better suited for capturing electronic structure characteristics relevant to band gap prediction. These findings highlight the complementary strengths of different architectures and establish CPGN as a competitive approach for materials property prediction, especially for band gap estimation. Due to the unavailability of the necessary implementation details and experimental setup, we were unable to reproduce the reported results in our own environment. Therefore, the comparison presented in Table~\ref{tab:extended_comparison_MPProject} is based exclusively on the performance metrics reported in the original publications. Although graph transformer architectures and pretrained language model-based approaches achieve good predictive performance, they often suffer from high computational complexity and resource requirements. The combination of GraphTransformer layers with GNN components increases memory usage, training time, and inference cost, while the quadratic complexity of self-attention limits scalability to large crystal structures. Pretrained models such as GIAT-PLM and SSFormer also depend on extensive pretraining datasets, specialized tokenization schemes, and substantial computational resources, making deployment more challenging. In contrast, CPGN provides a more computationally efficient alternative, requiring fewer parameters, simpler training procedures, and lower hardware demands while maintaining competitive performance. Notably, CPGN achieves the best band gap prediction accuracy among the compared methods, demonstrating that efficient GNN architectures can offer a favorable balance between predictive performance, scalability, and practical applicability for materials discovery.

\begin{table}[hbt!]
\centering
\scriptsize
\caption{Extended test results on the Materials Project dataset. The red and blue highlighted values represent the 1st best and 2nd best MAE, respectively.}
\label{tab:extended_comparison_MPProject}
\renewcommand{\arraystretch}{1.2}
\begin{tabular}{|c|cccccccc|}
\hline
\textbf{Metric} & \multicolumn{8}{c|}{\textbf{Mean Absolute Error (MAE)}} \\ \hline

Property &
\multicolumn{1}{c|}{Unit} &
\multicolumn{1}{c|}{MAD} &
\multicolumn{1}{c|}{GIN} &
\multicolumn{1}{c|}{SSformer} &
\multicolumn{1}{c|}{$GIN_{Barlow}$} &
\multicolumn{1}{c|}{GATGNN} &
\multicolumn{1}{c|}{GIAT-PLM} &
\textbf{CPGN} \\ \hline

Formation Energy ($E_f$) &
\multicolumn{1}{c|}{eV/atom} &
\multicolumn{1}{c|}{0.93} &
\multicolumn{1}{c|}{0.109} &
\multicolumn{1}{c|}{\textcolor{red}{0.039}} &
\multicolumn{1}{c|}{0.085} &
\multicolumn{1}{c|}{\textcolor{blue}{0.041}} &
\multicolumn{1}{c|}{0.042} &
0.060 \\ \hline

Band Gap ($E_g$) &
\multicolumn{1}{c|}{eV} &
\multicolumn{1}{c|}{1.35} &
\multicolumn{1}{c|}{0.601} &
\multicolumn{1}{c|}{0.360} &
\multicolumn{1}{c|}{0.337} &
\multicolumn{1}{c|}{0.323} &
\multicolumn{1}{c|}{\textcolor{blue}{0.314}} &
\textcolor{red}{0.292} \\ \hline

\textbf{Metric} & \multicolumn{8}{c|}{\textbf{MAD:MAE}} \\ \hline

Formation Energy ($E_f$) &
\multicolumn{1}{c|}{\multirow{2}{*}{ratio}} &
\multicolumn{1}{c|}{\multirow{2}{*}{--}} &
\multicolumn{1}{c|}{8.5} &
\multicolumn{1}{c|}{\textcolor{red}{23.84}} &
\multicolumn{1}{c|}{10.94} &
\multicolumn{1}{c|}{\textcolor{blue}{22.68}} &
\multicolumn{1}{c|}{22.14} &
15.5 \\ \cline{1-1} \cline{4-9}

Band Gap ($E_g$) &
\multicolumn{1}{c|}{} &
\multicolumn{1}{c|}{} &
\multicolumn{1}{c|}{2.25} &
\multicolumn{1}{c|}{3.75} &
\multicolumn{1}{c|}{4.01} &
\multicolumn{1}{c|}{\textcolor{blue}{4.17}} &
\multicolumn{1}{c|}{4.29} &
\textcolor{red}{4.6} \\ \hline

\end{tabular}
\end{table}

\section{Conclusions}
\label{conclusion}

This paper presents the Coordination Polyhedron Graph Network (CPGN), a novel multi-task graph neural network framework for crystal and molecular property prediction that incorporates three complementary graph representations: an atom graph, a line graph, and a coordination polyhedron graph. By integrating Voronoi-derived local coordination environments with bidirectional cross-attention between atom and polyhedron embeddings, CPGN captures physically grounded structural priors that are not fully represented in conventional atom-centric graph neural networks. Extensive evaluations across multiple benchmark datasets demonstrate that CPGN achieves competitive and, in several cases, superior predictive performance. On the Materials Project dataset, CPGN attains a formation energy MAE of $0.060~\text{eV/atom}$ and the lowest band gap MAE of $0.292~\text{eV}$. On the JARVIS-DFT benchmark, it outperforms CGCNN and ALIGNN across several properties, including formation energy ($0.092~\text{eV/atom}$), optical band gap ($0.227~\text{eV}$), bulk modulus ($14.01~\text{GPa}$), shear modulus ($11.28~\text{GPa}$), and thermodynamic stability ($0.088~\text{eV/atom}$). On the QM9 dataset, CPGN achieves competitive MAEs for HOMO ($0.030~\text{eV}$), LUMO ($0.023~\text{eV}$), and the HOMO--LUMO gap ($0.038~\text{eV}$), outperforming MEGNet and SchNet while approaching DimeNet++ and ALIGNN. The slightly higher error in ZPVE prediction highlights the challenges of modeling vibrational properties without explicit normal-mode information. Despite these promising results, several limitations remain. The coordination polyhedron graph relies on Voronoi tessellation, which may be sensitive to structural noise or imperfect geometries. The auxiliary loss weighting ($\lambda_{\text{aux}} = 0.1$) is fixed, potentially limiting optimal task balancing in the multi-task setting. Additionally, the observed train--validation gap on the Materials Project dataset suggests that the increased model capacity may require larger datasets or stronger regularization. Furthermore, CPGN has not yet been evaluated on low-dimensional or disordered materials. Future work will explore adaptive formulations of structural and spectral parameters, including learnable Voronoi and frequency cutoff mechanisms. Extending CPGN to large-scale pre-training across combined molecular and crystalline datasets may further enhance generalization and transferability. Finally, applying CPGN to downstream tasks such as materials screening, inverse design, and anomaly detection represents a promising direction for real-world deployment in high-throughput materials discovery pipelines. Future work will introduce a binary classification head trained with weighted cross-entropy to predict stability, and evaluate performance on the Matbench Discovery benchmark~\cite{dunn2020benchmarking} against models such as CGCNN, MEGNet, ALIGNN, M3GNet, and CHGNet. Given CPGN’s strong performance on geometry-sensitive properties like band gaps, it is expected that its explicit polyhedral inductive bias will improve stability prediction, particularly for materials where instability arises from local structural distortions rather than composition alone.

\section*{Acknowledgements}
\subsection*{Funding:} This work was partially supported by Project funding (KAW 2020.0033), the Wallenberg AI, Autonomous Systems and Software Program (WASP), and the Wallenberg Initiative Materials Science for Sustainability (WISE), all funded by the Knut and Alice Wallenberg Foundation. We would also like to thank Vinnova and the BioInnovation program.

\subsection*{Author Contributions}
\textbf{Sanjay Chakraborty:} Conceptualization, Methodology, Software, Data curation, Writing- Original draft preparation, Visualization, Investigation. 
\textbf{Fredrik Heintz:} Supervision, Validation, Reviewing and Editing. 

\subsection*{Code and Data Availability}
The source code for the proposed method is publicly available at: 

https://github.com/sanjaylopa22/CPGN-Coordination-Polyhedron-Graph-Neural-Network

\subsection*{Conflicts of Interest}
The authors declare no competing interests.

\bibliographystyle{unsrt} 
\bibliography{mybibfile}

\end{document}